\documentclass[pra,aps,twocolumn,showpacs]{revtex4-1}
\usepackage{amssymb}
\usepackage{amsmath}
\usepackage{amsfonts}
\usepackage{graphicx}
\usepackage{color}
\usepackage{bm}
\usepackage[sort&compress]{natbib}

\renewcommand{\Re}{\operatorname{Re}}
\renewcommand{\Im}{\operatorname{Im}}

\begin{document}

\title{Interference in $\omega -2\omega $ atomic ionization within the
strong-field approximation: Beyond the perturbative regime}
\author{Diego G. Arb\'{o}$^{1,2}$, Sebasti\'{a}n D. L\'{o}pez$^{1},$}
\affiliation{$^{1}$ Institute for Astronomy and Space Physics - IAFE (CONICET-UBA), CC
67, Suc. 28, C1428ZAA, Buenos Aires, Argentina.}
\affiliation{$^{2}$Universidad de Buenos Aires, Facultad de Ciencias Exactas y Naturales
y Ciclo B\'{a}sico Com\'{u}n, Buenos Aires, Argentina.}
\date{\today }

\begin{abstract}
We analyze interference processes in atomic ionization induced by a
two-color laser with fundamental frequency $\omega $ and its second harmonic 
$2\omega $. The interplay between inter- and intracycle interference
processes give rise to multiphoton peaks which can be named as main or ATI
peaks and sidebands, in analogy to the well-known RABBIT (reconstruction of
attosecond harmonic beating by interference of two-photon transitions). We
use the saddle point approximation (SPA) to extract the complex ionization
times of the interfering electron trajectories. Changing the relative phase
between the two colors, the doubly differential momentum distribution of
emitted electrons can be controlled. We study the dependence of the electron
emission as a function of the relative phase between the $\omega $ and $%
2\omega $ fields within the strong field approximation (SFA) but beyond the
perturbative regime. We focus on the extraction of the phase delays
accounting the electron forward emission in the direction of the polarized
electric fields. We characterize the time delays in the emission of
electrons for visible frequency of the pump and its first harmonic as a
probe [Ti:Sapphire laser ($800$ nm) together with the first harmonic ($400$
nm)] for a typical $\omega -2\omega $ configuration for argon ionization. We
find excellent agreement between our SPA results and the corresponding SFA
(without any further approximation) and also with previous perturbative
theories.
\end{abstract}

\date{\today }
\maketitle

\section{Introduction}

Photoionization is defined as the detachment of one or multiple electrons
from a system, such as an atomic or molecular ion, a cluster, or a solid,
due to the electromagnetic force exerted by external radiation. Two
different pictures has been widely used to depict the photoionization
process. On one hand, the exiting electron tunnels through the barrier
formed by the parent structure and the external field when the time that the
electron travels through the barrier is longer than the time variation of
the external field (situation known as the tunneling regime). And on the
other hand, when the variation of the external field is very rapid,
tunneling is not possible and photoionization proceeds through the
absorption of one or several photons allowing the target gain energy up to
the continuum near threshold or even higher (situation known as the
multiphoton regime) \cite{Keldysh64, Keldysh65, Bergues07}. First
experiments used rather weak lasers, for which, the ionization processes
were deep in the multiphoton regime \cite{Agostini79}. Theories accompanying
these experiments were firstly perturbative \cite{Faisal73a, Faisal73b} but,
as irradiance of laser beams grew, these perturbation theories became
obsolete and a new paradigm was necessary \cite{Corkum94, Lewenstein95,
Ivanov95}. Based on rather simple pictures of the photoionization processes
like the simpleman's model (SMM) or the strong field approximation (SFA),
interference structures in photoelectron spectra have been identified as a
diffraction pattern from a time grating\ composed of inter- and intracycle
interferences \cite{Bivona08,Arbo10a,Arbo10b,Arbo12}. Whereas the intercycle
interferences give rise to multiphoton peaks, intracycle interferences lead
to a modulation of the ATI spectrum offering information on the subcycle
ionization dynamics \cite{Xie12,Arbo10a,Arbo14a}.

Metrology of atomic processes became accessible through pump-probe
techniques such as attosecond streaking \cite{Itatani02,
Goulielmakis04,Goulielmakis07} and RABBIT (reconstruction of attosecond
harmonic beating by interference of two-photon transitions) \cite%
{Veniard95,Veniard96,Paul01}. These techniques involve lasers of at least
two very different frequencies: Electrons are emitted due to the absorption
of an XUV field and probed by a moderately intense field in the
near-infrared region of the electromagnetic spectrum. In this way, it was
possible to measure phase shifts compatible with attosecond delays for noble
gas atoms \cite{Schultze10,Klunder11,Guenot12, Guenot14}. For attosecond
streaking, the oscillating probe field moves classically the electron
previously freed by the pump field, producing gains and losses of its final
kinetic energy \cite{Itatani02, Fuchs20}. In RABBIT, ionization is given by
two consecutive high harmonics fields of odd parity followed by absorption
or emission of a photon with the fundamental frequency. By analyzing the
interference between these two ionization paths, the intrinsic phase shifts
in above threshold ionization (ATI) spectra could be extracted by means of
second-order perturbative calculations \cite{Dahlstrom13} or by analyzing
the asymptotic behavior of the scattered electron wave packet from
numerically exact solutions of the time-dependent Schr\"{o}dinger equation
(TDSE) \cite{Nagele12,Pazourek12,Kheifets13,Feist14, Su13,Pazourek15,Boll16}.

Strong-field ionization by laser fields with commensurate frequencies and
well-defined relative phase permits the tune and control of the emission
process \cite{Schumacher94,Arbo15,Ehlotzky01,Xie12,Arbo14a}. ATI by laser
pulses using the fundamental component and one of its harmonics were
investigated \cite{Ehlotzky01,Schumacher94,Muller90} and applied for
controlled ionization \cite{Thompson97,Sheehy95,Ohmura04}, dichroism \cite%
{Fifirig03,Cionga03}, orientation of molecules \cite{De09}, and control of
interference fringes in the electron momentum distribution \cite%
{Xie12,Arbo14a}. The temporal shape of the two-color field is determined by
the intensities of the two components and their relative phase. Coherent
phase control refers to the manipulations of some physical processes through
the relative phase\ \cite{Ehlotzky01}. The concept of phase shifts and time
delays in RABBIT has been extended by Zipp \textit{et al.} \cite{Zipp14} for
two-color ($\omega -2\omega $) lasers with controlled relative phase. Very
recently, we have theoretically explored the extraction of $\omega -2\omega $
phase delays by means of the \textit{ab initio} solution of the TDSE and
also through the development of a perturbation theory \cite{Lopez21}.

In this work, we developed a non-perturbation theory of the electronic
photoemission process in atomic argon due to a two-color ($\omega -2\omega $%
) linearly polarized short laser pulse in the multiphoton regime. In Sec. %
\ref{theory}, we introduce the general theory based on the saddle point
approximation (SPA) to calculate the ionization time of each interfering
electron trajectory, firstly analyzing one-color ionization and then the $%
\omega -2\omega $ setup. Different interference structures of the doubly
differential momentum distribution are analyzed. We focus on the extraction
of the phase shifts using directional emission in the forward direction. We
show that our SPA results are in excellent agreement with the results of the
SFA (without any further approximation) and also the perturbation theories
in the literature \cite{Bertolino21,Lopez21}. We make our final remarks in
Sec. \ref{conclusions}. Atomic units ($e=\hbar =m_{e}=1$ a.u.) are used
throughout unless stated otherwise.

\section{Non-perturbative strong-field approximation}

\label{theory}

In general, ionization of an atomic system by a linearly polarized laser
pulse can be considered in the single-active-electron approximation. The
TDSE then reads 
\begin{equation}
i\frac{\partial }{\partial t}\left\vert \psi (t)\right\rangle =\left[
H_{0}+H_{\text{int}}(t)\right] \left\vert \psi (t)\right\rangle ,
\label{TDSE}
\end{equation}%
where $H_{0}=\vec{p}^{2}/2+V(r)$ is the time-independent atomic Hamiltonian,
whose first term corresponds to the electron kinetic energy and its second
term to the electron-core Coulomb interaction. In Eq. (\ref{TDSE}), $H_{%
\text{int}}(t)$ corresponds to the interaction hamiltonian between the
atomic system and the external radiation field. Because of the presence of
the external laser field, the electron initially bound in an atomic state $%
|\phi _{i}\rangle $ can either remain in the same state, be excited to
another atomic bound state, or be emitted to a final continuum state $|\phi
_{f}\rangle $ with final momentum $\vec{k}$ and energy $E=k^{2}/2$. In the
latter case, we call the process photoionization and the transition
amplitude within the time-dependent distorted wave theory in the prior form
is expressed as \cite{Macri03, Arbo08a}%
\begin{equation}
T_{\mathrm{if}}=-i\int_{-\infty }^{+\infty }dt\,\langle \chi _{f}^{-}(\vec{r}%
,t)|H_{\text{int}}(\vec{r},t)|\phi _{i}(\vec{r},t)\rangle ,  \label{Tif}
\end{equation}%
where $\phi _{i}(\vec{r},t)=\varphi _{i}(\vec{r})\,e^{iI_{p}t}$ is the
initial atomic state with ionization potential $I_{p}$ and $\chi _{f}^{-}(%
\vec{r},t)$ is the distorted final state. Eq. (\ref{Tif}) is exact as far as
the final channel $\chi _{f}^{-}(\vec{r},t)$ is the exact solution of Eq. (%
\ref{TDSE}), within the dipole approximation. Throughout this paper, we will
be considering linearly polarized laser fields (in the $\hat{z}$ direction).

Several degrees of approximation have been considered in the literature to
solve Eq. (\ref{Tif}). The widest known one is the strong field
approximation (SFA), which neglects the Coulomb distortion (in the final
channel) produced on the ejected-electron state due to its interaction with
the residual ion and discard the influence of the laser field in the initial
ground state~and the depletion of the ground state \cite%
{Lewenstein94,Lewenstein95}. The SFA consists in approximating the distorted
final state with the solution of the TDSE for a free electron in an
electromagnetic field, namely, a Volkov function \cite{Volkov35}, i.e., $%
\chi _{f}^{-}(\vec{r},t)=\chi _{f}^{V}(\vec{r},t)$, where 
\begin{eqnarray}
\chi _{f}^{V}(\vec{r},t) &=&\frac{1}{(2\pi )^{3/2}}\exp \{i[\vec{k}+\vec{A}%
(t)]\cdot \vec{r}\}  \notag \\
&\times &\exp \left\{ \frac{i}{2}\int_{t}^{\infty }[\vec{k}+\vec{A}%
(t^{\prime })]^{2}dt^{\prime }\right\}  \label{Volkov}
\end{eqnarray}%
the vector potential due to the total external field is defined as $\vec{A}%
(t)=-\int_{-\infty }^{t}dt^{\prime }\vec{F}(t^{\prime }),$ and $\vec{F}(t)$
denotes the external laser field. The final Volkov function in Eq. (\ref%
{Volkov}) is calculated within the length gauge, i.e., $H_{\text{int}}(\vec{r%
},t)=\vec{F}(t)\cdot \vec{r}$.

Therefore, $T$-matrix in Eq. (\ref{Tif}) can be written as%
\begin{equation}
T_{\mathrm{if}}=\int_{-\infty }^{+\infty }\,\ell (t)\ e^{iS(t)}\,\,dt,
\label{Tm}
\end{equation}

where 
\begin{eqnarray}
\ell (t) &=&-i\vec{F}(t)\cdot \vec{d}\left[ \vec{k}+\vec{A}(t)\right]  \notag
\\
&&\mathrm{and}  \notag \\
S(t) &=&-\int_{t}^{\infty }dt^{\prime }\left\{ \frac{\left[ \vec{k}+\vec{A}%
(t^{\prime })\right] ^{2}}{2}+I_{p}\right\}  \label{action}
\end{eqnarray}%
with the dipole transition moment defined as $\vec{d}(\vec{v})=(2\pi
)^{-3/2}\langle e^{i\vec{v}\cdot \vec{r}}|\vec{r}|\varphi _{i}(\vec{r}%
)\rangle $, and $S(t)$ is the Volkov action.

We assume that the pump field is composed of $2N$ optical cycles each of
duration $T=2\pi /\omega $. Then,

\begin{eqnarray}
T_{\mathrm{if}} &=&\int_{0}^{NT}\,\ell (t)e^{iS(t)}\,\,dt  \notag \\
&=&\sum_{j=0}^{N-1}\int_{jT}^{(j+1)T}\ell (t)e^{iS(t)}dt.  \label{Tm1}
\end{eqnarray}%
We consider now a general electric field (and vector potential) with a
smooth envelope with a central flat-top region where both $\vec{F}(t)$ and $%
\vec{A}(t)$ are oscillating with period $T$. From Eq. (\ref{action}), it is
straightforward to realize that $S(t)-at$ is a time-oscillating function
with the same period of the laser field and vector potential, 
\begin{equation}
S(t+jT)=S(t)+ajT,  \label{S-periodic}
\end{equation}%
where 
\begin{equation}
a=\frac{k^{2}}{2}+I_{p}+U_{p},  \label{a}
\end{equation}%
and $U_{p}=\int_{t}^{t+T}dt^{\prime }A(t^{\prime })^{2}.$ In light of the
periodicity properties of the action in Eq. (\ref{S-periodic}) and that $%
\ell (t+jT)=\ell (t)$, the transition matrix $T_{\mathrm{if}}$ in Eq. (\ref%
{Tm}) can be written in terms of the contribution of the first fundamental
cycle or unit cell \cite{Arbo12,DellaPicca20}, 
\begin{eqnarray}
T_{\mathrm{if}} &=&\sum_{j=0}^{N-1}\int_{jT}^{(j+1)T}\ell
(t+jT)e^{iS(t+jT)}dt  \notag \\
&=&\sum_{j=0}^{N-1}e^{iajT}\int_{0}^{T}\ell (t)e^{iS(t)}dt  \notag \\
&=&\frac{\sin {(aTN/2)}}{\sin {(aT/2)}}\,e^{(iaT(2N-1)/2)}I(\vec{k}).
\label{Tm3}
\end{eqnarray}

From the absolute value of the transition matrix we can extract
probabilistic information, like the doubly differential momentum
distribution or the angle resolved photoelectron spectrum. Because of the
azimuthal symmetry, the electron distribution can be expressed in terms of
only two physical magnitudes, i.e., the final electron momentum parallel $%
k_{z}$ and transversal $k_{\rho }$ to the field polarization direction or,
alternatively, the final kinetic energy $E$ and the final polar emission
angle $\theta $: 
\begin{equation}
\left\vert T_{\mathrm{if}}\right\vert ^{2}=\frac{dP}{2\pi k_{\rho }dk_{\rho
}dk_{z}}=\frac{dP}{2\pi \sqrt{2E}\ dE\ d(\cos \theta )}.  \label{dPdE}
\end{equation}

The factor $I(\vec{k})=\int_{0}^{T}\ell (t)e^{iS(t^{\prime })}dt^{\prime }$
in Eq. (\ref{Tm3}) corresponds to the contribution into one optical cycle of
the $\omega $ field and $|I(\vec{k})|^{2}$ is known in the literature as the 
\emph{intracycle} contribution to the ionization probability \cite%
{Arbo08a,Arbo08b,DellaPicca20}. Thus, the photoelectron spectrum (PES) can
be expressed as a product of the \emph{intracycle} factor $|I(\vec{k})|^{2}$
and the \emph{intercycle} factor $\left( \sin {(aTN/2)}/\sin {(aT/2)}\right)
^{2}$, being the latter the result of the phase interference arising from
the $N$ different optical cycles of the field \cite{Arbo10a,Arbo10b,Arbo12}.
We want to point out that Eq. (\ref{Tm3}) is a mere consequence of the
periodicity of the transition matrix with no further approximations, except
for a flat-top pulse \cite{DellaPicca20}.

Finite maxima are reached at the zeroes of the denominator of the intercycle
factor $\left( \sin {(aTN/2)/}\sin {(aT/2)}\right) ^{2}$, i.e., the energy
values satisfying $aT/2=n\pi $, since the numerator also cancels out at
these points. Such maxima are recognized as the multiphoton peaks of the
PES. They occur when 
\begin{equation}
E_{n}=n\omega -I_{p}-U_{p},  \label{ATI1}
\end{equation}%
where we have used Eq. (\ref{a}). In fact, when $N\rightarrow \infty $, the
intercycle factor becomes a series of delta functions, i.e., $\sum_{n}\delta
(E-E_{n})$, satisfying the conservation of energy. Instead, for finite pulse
durations $\tau =NT$ (composed of $N$ cycles), each multiphoton peak has a
width $\Delta E\sim 2\pi /NT$, fulfilling the uncertainty relation $\Delta
E\tau \sim 2\pi $.

The intracycle amplitude $I(\vec{k})=\int_{0}^{T}\ell (t)e^{iS(t^{\prime
})}dt^{\prime }$ in Eq. (\ref{Tm3}) can be calculated either numerically
(SFA) or within the saddle point approximation (SPA). In the latter, the
intracycle amplitude can be regarded as a superposition of all electron
trajectories within any optical cycle or unit cell with final momentum $\vec{%
k}$ 
\begin{equation}
I(\vec{k})\simeq \sum_{\beta }\ell (t_{\beta })\frac{e^{iS(t_{\beta })}}{%
\left\vert \ddot{S}(t_{\beta })\right\vert ^{1/2}},  \label{I1}
\end{equation}%
each starting at a complex ionization times $t_{\beta }$ fulfilling the
saddle equation $\dot{S}(t_{\beta })=0$ (where the dot denotes the time
derivative), i.e., 
\begin{equation}
\frac{\left[ \vec{k}+\vec{A}(t_{\beta })\right] ^{2}}{2}+I_{p}=0.
\label{saddle}
\end{equation}%
In general, solutions of Eq. (\ref{saddle}) come in pairs ($t_{\beta
},t_{\beta }^{\ast }$), where the star means complex conjugate. From each
couple, we select only the solution with positive imaginary part to avoid
spurious exponential growth of probabilities and only keep exponential
decays when calculating $\exp \left[ iS(t_{\beta })\right] =\exp \left\{ i%
\Re\left[ S(t_{\beta })\right] \right\} \exp \left\{ -\Im\left[
S(t_{\beta })\right] \right\} $ in Eq. (\ref{I1}). The SMM considers real
ionization times by neglecting $I_{p}$ and the perpendicular momentum
reducing Eq. (\ref{saddle}) to $k_{z}+A(t_{\beta })=0.$

\subsection{One color photoionization}

For the case of atomic photoionization by a one color field $\vec{F}%
(t)=Ff(t)\cos (\omega t+\phi )\ \hat{z}$ with $f(t)$ a smooth function
between $0$ and $1$ mimicking the pulse envelope, $\hat{z}$ the polarization
direction, and $F$ the field strength, the action can be calculated from Eq.
(\ref{action}) as 
\begin{equation}
S_{0}(t)=at+b\cos (\omega t)+c\sin (2\omega t),  \label{action-1c}
\end{equation}%
where $a$ is given by Eq. (\ref{a}), $b=F/\omega ^{2}\hat{z}\cdot \vec{k},$ $%
c=-U_{p}/2\omega $, and the ponderomotive energy is $U_{p}=(F/(2\omega
))^{2} $. We denote $S_{0}$ the one-color action just to distinguish it from
the two-color action in the next subsection.

From Eq. (\ref{saddle}), two ionization times can be analytically calculated
with the following expressions

\begin{eqnarray}
t_{1} &=&\frac{1}{\omega }\sin ^{-1}\left[ \frac{\omega }{F}\left( k_{z}+i%
\sqrt{2I_{p}+k_{\rho }^{2}}\right) \right] ,  \notag \\
t_{2} &=&\frac{\pi }{\omega }-t_{1}^{\ast },  \label{ts0p}
\end{eqnarray}%
for $k_{z}\geq 0$ and,%
\begin{eqnarray}
t_{1} &=&\frac{\pi }{\omega }+\frac{1}{\omega }\sin ^{-1}\left[ \frac{\omega 
}{F}\left( k_{z}+i\sqrt{2I_{p}+k_{\rho }^{2}}\right) \right] ,  \notag \\
t_{2} &=&\frac{3\pi }{\omega }-t_{1}^{\ast },  \label{ts0m}
\end{eqnarray}%
for $k_{z}\leq 0.$

\begin{figure}[tbp]
	\includegraphics[width=6cm]{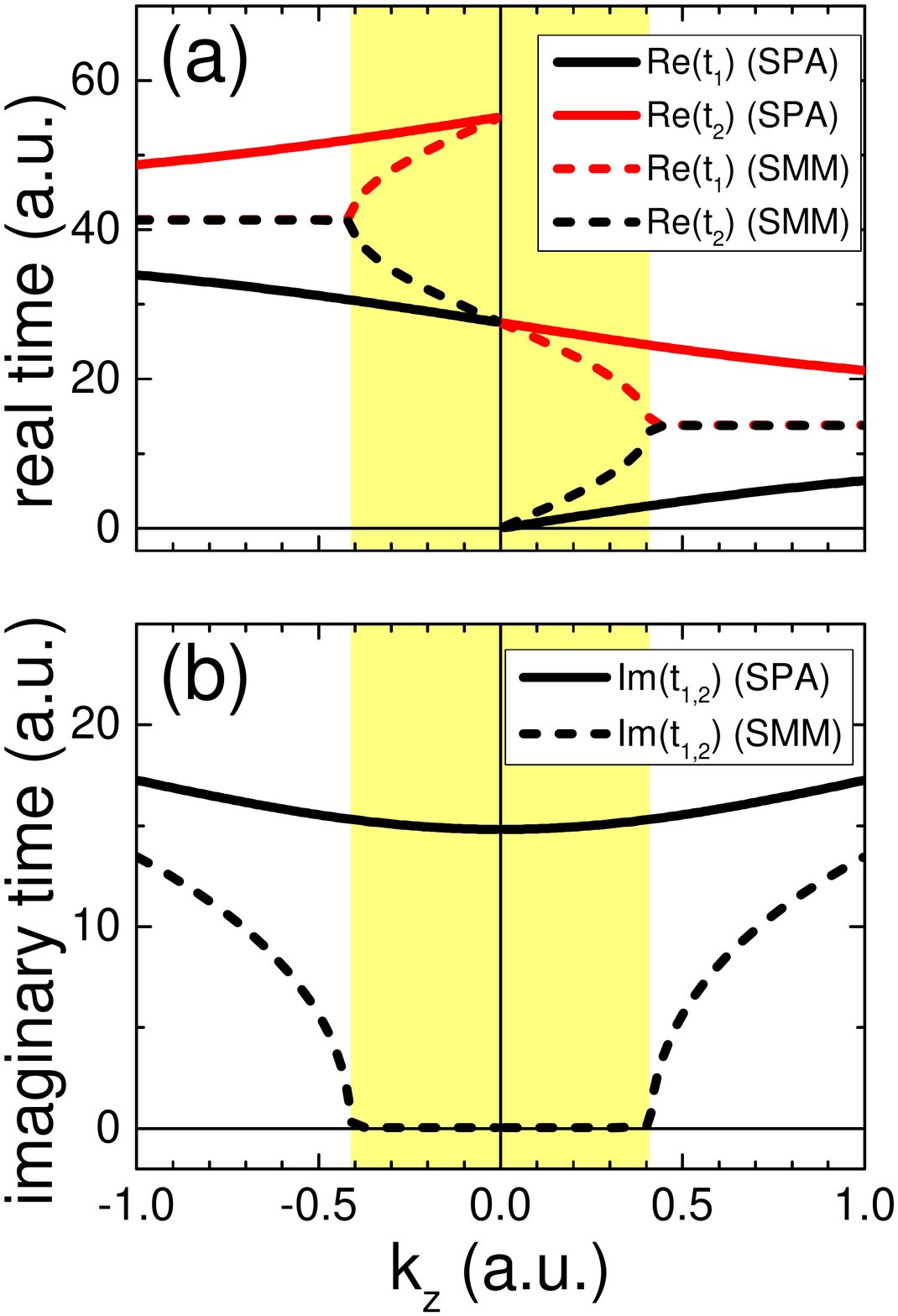}
	\caption{Complex saddle times $t_{1}$ and $t_{2}$ as a function of the longitudinal momentum
	$k_{z}$ for $k_{\bot }=0$ for one-color ionization. In solid (dash) lines the results of
the SPA (SMM, including an ionization potential of $10^{-6}$ to generate the
imaginary part of the saddle time). (a) In black $\Re(t_{1})$ and in
red $\Re(t_{2})$. (b) In solid (dash) lines the results of the SPA
(SMM) $\Im(t_{1})=\Im(t_{2})$. The yellow region in the $k_{z}$
domain corresponds to the allowed classical region in the SMM. In the
classical-forbiden region, the imaginary part of the SMM saddle times
increases.}
\label{saddle-time-1c}
\end{figure}

We show the real part of $t_{1}$ and $t_{2}$ in Fig. \ref{saddle-time-1c}a and the imaginary
parts in Fig. \ref{saddle-time-1c}b for an electric field of strength $F=0.0469$ a.u. and
frequency $\omega =0.114,$ for the special case that $k_{\bot }=0$ (forward
and backward emission). The SMM times are drawn in dash line and were
calculated by including a tiny ionization potential $I_{p}=10^{-6}.$ The
shaded region of longitudinal momentum $|k_{z}|\leq 2\sqrt{U_{p}}$
corresponds to the classical accessible region for the electron according to
the SMM. Inside this region $\Im(t_{1})=\Im(t_{2})=0$ while
outside, $\Im(t_{1})=\Im(t_{2})>0.$ In turn, for the SPA, we
have used an ionization potential $I_{p}=0.58$ a.u. (corresponding to the
ground state of atomic argon).

\begin{figure}[tbp]
	\includegraphics[width=8cm]{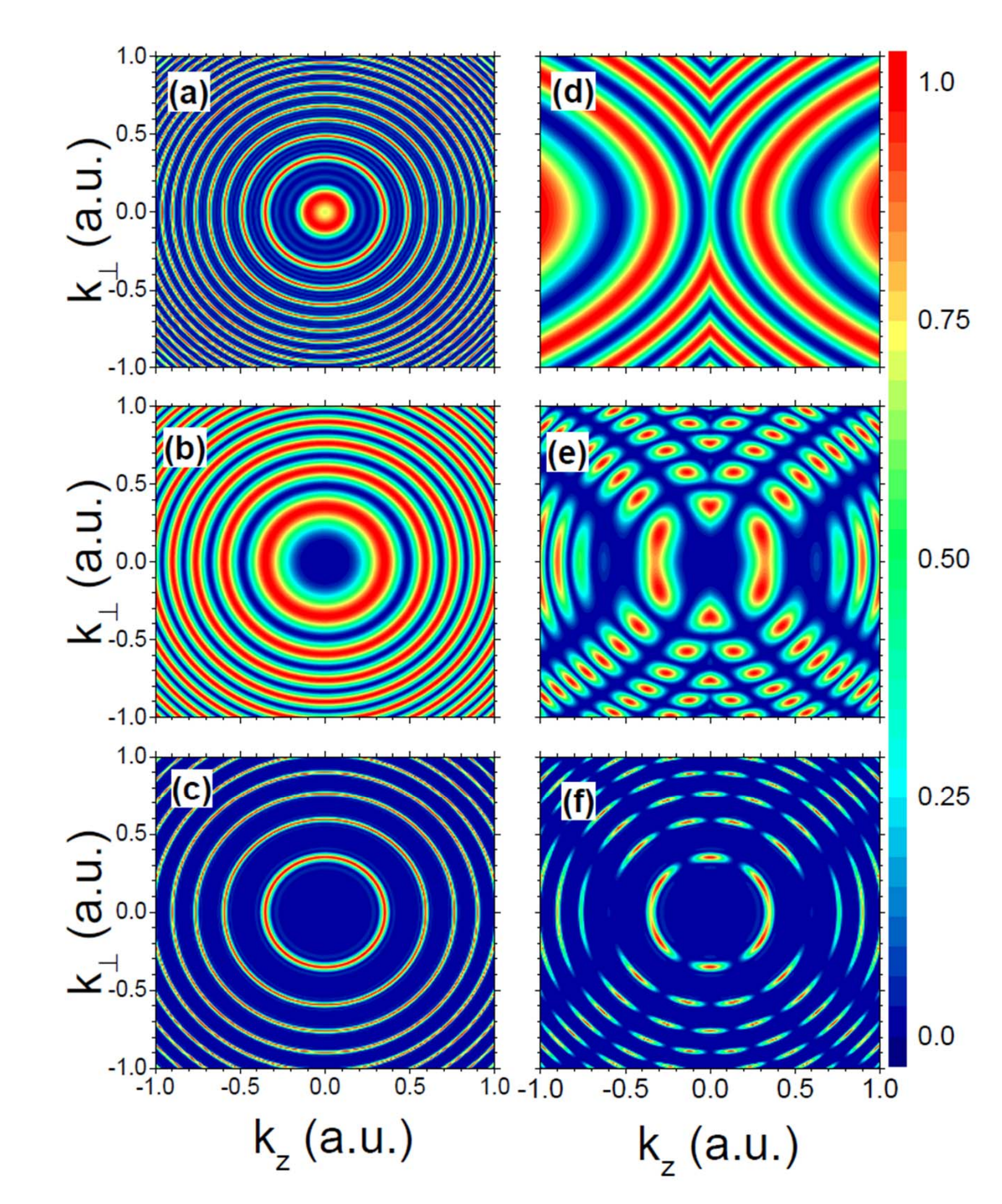}
	\caption{Doubly differential momentum
distribution as a function of the longitudinal momentum $k_{z}$ and the
perpendicular momentum $k_{\bot }$ for one-color ($2\omega $) ionization.
(a) Intercycle factor considering $4\pi /\omega $ periodicity generating ATI
and SB rings, (b) Intercycle factor considering $2\pi /\omega $ periodicity
generating only ATI rings (of double width), (c) Multiplication of (a) and
(b), the ATI rings are visible but the SB rings disappeared since they
coincide with the minima of the distribution in (b). (d) corresponds to the
intracycle interference, (e) to the multiplication of distributions in (d)
and (b), and in (f) the total momentum distribution is displayed
(multiplication of (c) and (d)).}
\label{1c-kzkrho}
\end{figure}

In Fig. \ref{1c-kzkrho}a we show the intercycle factor $\left( \sin {(aTN/2)/}\sin {(aT/2)}%
\right) ^{2}$ with $N=4$ as a function of the longitudinal momentum $k_{z}$
and the perpendicular momentum $k_{\bot }$ (whose magnitude is equal to $%
k_{\rho }.$ Isotropic rings are observed at radii $k_{n}=\sqrt{2E_{n}},$
where $E_{n}$ are given by Eq. (\ref{ATI1}), corresponding to the absorption
of $n$ photons. The intracycle factor is displayed in Fig. \ref{1c-kzkrho}d exhibiting a
strong angular dependence. The total doubly differential momentum
distribution is proportional to the multiplication of the intercycle and
intracycle factors and is displayed in Fig 2f. In this paper we omit in the
calculations the factor $l(t)$ containing the atomic dipole moment just to
focus on interference aspects of the photoionization processes.

\subsection{Interference in ($\protect\omega -2\protect\omega $)
photoionization}

The main goal of this work is to extend the well-known interference
structures of electron photoemission in one-color atomic photoionization to
the case of two colors, where one main frequency (harmonic) doubles the
other (fundamental), i.e., $\omega -2\omega $ photoionization emission,
within the SFA. If the fundamental intensity is very low compared to the
intensity of its second harmonic, some connection between $\omega -2\omega $
ionization and RABBIT can be speculated and analyzed showing some
similarities and some differences \cite{Zipp14,Bertolino21,Lopez21}. In the
RABBIT jargon for even $n$ in Eq. (\ref{ATI1}), the energy maxima are named
ATI peaks and for odd $n$ they are named sidebands, but although in our
context that denomination of the multiphoton peaks is arbitrary, we will
maintain it for clarity. In general $n=n_{2\omega }+$ $n_{\omega }$ denotes
the absorption (emission) of a $n_{2\omega }$ number of $2\omega $ photons
and absorption (emission) and a $n_{\omega }$ number of $\omega $ photons
for positive (negative) $n_{2\omega }$ and $n_{\omega }$ values. We consider
the two-color electric field of the form

\begin{figure}[tbp]
	\includegraphics[width=8cm]{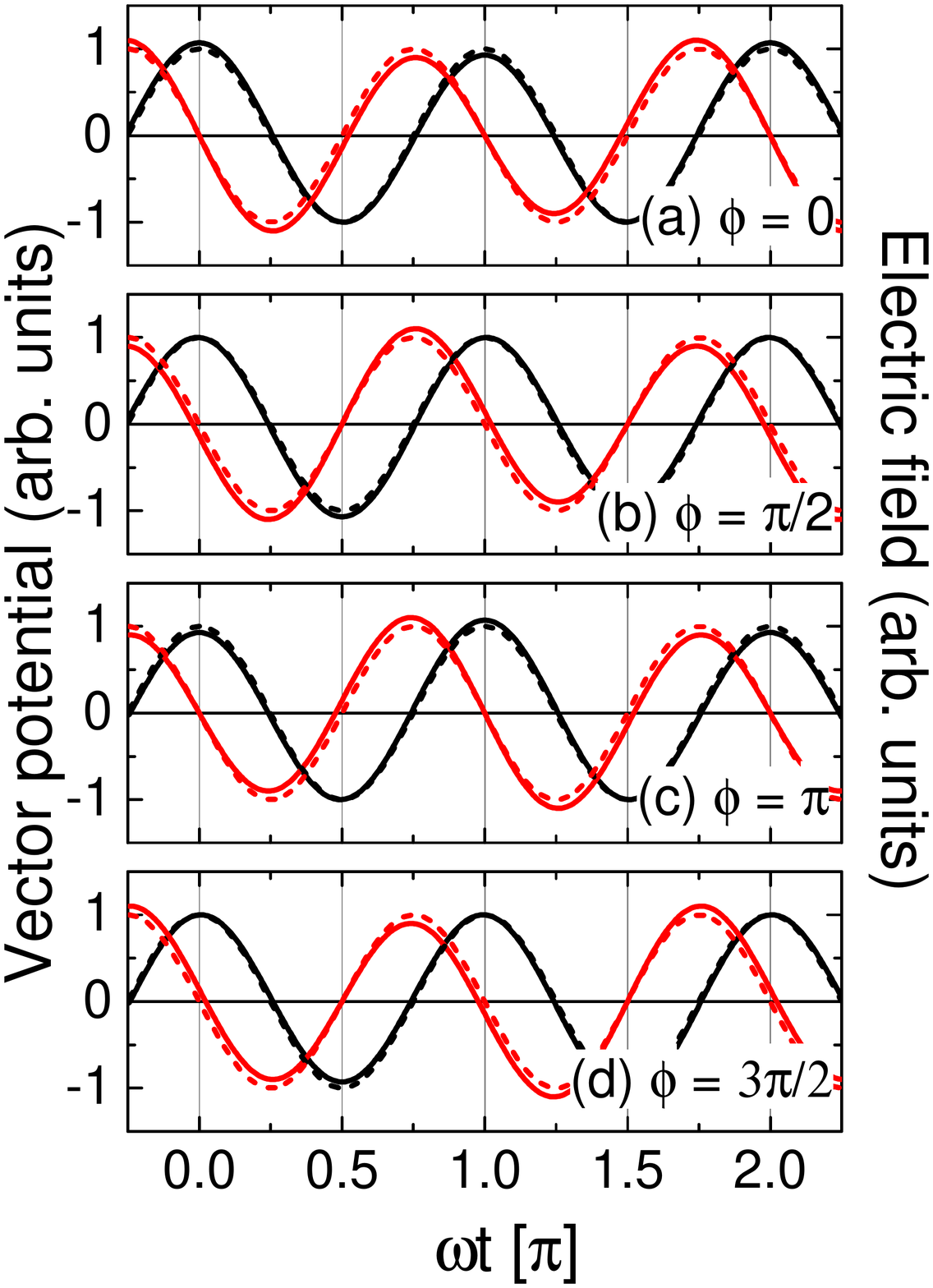}
	\caption{Vector potential (red) and
electric field (black) along the polarization axis as a function of time. In
solid line the two-color version and in dash line the one-color version
(switching off the $\omega $ terms in Eqs. (\ref{field2}) and (\ref{Avector}%
). (a) for $\phi =0,$ (b) for $\phi =\pi /2,$ (c) for $\phi =\pi ,$ and (d)
for $\phi =3\pi /2.$ Vector potentials and electric fields are normalized
with respect to their respective one-color version.}
\label{fields-2c}
\end{figure}

\begin{equation}
\vec{F}(t)=f(t)\left[ F_{2\omega }\cos \left( 2\omega t\right) +F_{\omega
}\cos (\omega t+\phi )\right] \ \hat{z},  \label{field2}
\end{equation}%
with $\phi $ the relative phase of the second harmonic with respect to the
fundamental laser field, $f(t)$ is a smooth function between $0$ and $1$
mimicking the pulse envelope, $\hat{z}$ is the polarization direction of
both fields, and $F_{2\omega }$ and $F_{\omega }$ are the field strengths of
the second harmonic and fundamental frequency, respectively. For a long
pulse with adiabatic switch on and off the vector potential can be written
in its central part $f(t)\simeq 1$), as 
\begin{equation}
\vec{A}(t)=-f(t)\left[ \frac{F_{2\omega }}{2\omega }\sin (2\omega t)+\frac{%
F_{\omega }}{\omega }\sin (\omega t+\phi )\right] \hat{z},  \label{Avector}
\end{equation}%
giving rise to a periodicity property of the vector potential and the
electric field, i.e., $\vec{A}(t)=\vec{A}(t+2j\pi /\omega )$ and $\vec{F}(t)=%
\vec{F}(t+2j\pi /\omega ),$ with $j$ any integer number provided that $%
f(t+2j\pi /\omega )=1.$ For our calculations we use the same parameters as
in Ref. \cite{Zipp14} with $F_{2\omega }=0.0469$ a.u. ($I_{2\omega }=8\times
10^{13}$ W/cm$^{2}$) and $F_{\omega }=0.00332$ a.u. ($I_{\omega }=4\times
10^{11}$ W/cm$^{2}$). In Fig. \ref{fields-2c} we show the electric field in black and the
vector potential in red as a function of time. We have normalized both
fields to the one-color case ($F_{\omega }=0$) displayed in dotted lines. As
the one-color fields show obviously invariant under changes of the relative
phase, there are small but appreciable changes in the two-color fields. For
example, the one-color fields are $\pi /\omega $-periodic, whereas the
two-color fields are $2\pi /\omega $-periodic. As shown for the one-color
case \cite{Arbo10a,Arbo10b}, the correct way to choose the unit cell
corresponding to one optical cycle of the field is from a root of the vector
potential. For the one-color case this corresponds to, for example, $t=0,$
whereas this value changes for the two-color case as a function of the
relative phase $\phi .$ For example, considering the roots of the vector
potential closest to the origin as the left border of the unit cells, they
are determined by $t\in \left[ 0,2\pi /\omega \right] $ ($\phi =0$ in Fig.
3a), $t\in \left[ -1.243,-1.243+2\pi /\omega \right] $ (in a.u. for $\phi
=\pi /2$ in Fig 3b), $t\in \left[ 0,2\pi /\omega \right] $ ($\phi =\pi $ in
Fig. \ref{fields-2c}c), and $t\in \left[ 1.243,1.243+2\pi /\omega \right] $ (in a.u. for $%
\phi =3/2$ in Fig. \ref{fields-2c}d).

Under the assumption of adiabatic switch on and off, the action in Eq. (\ref%
{action}) can be analytically calculated (in the central region where $%
f(t)=1 $) as

\begin{eqnarray}
S(t) &=&at+b\cos (2\omega t)+c\sin (4\omega t)+d\cos (\omega t+\phi )
\label{action2} \\
&&+e\sin (\omega t-\phi )+f\sin (2\omega t+2\phi )+g\sin (3\omega t+\phi ), 
\notag
\end{eqnarray}%
where $a$ is given by Eq. (\ref{a}), and 
\begin{eqnarray}
b &=&\frac{F_{2\omega }}{4\omega ^{2}}\hat{z}\cdot \vec{k},  \label{abc} \\
c &=&-\frac{U_{p,2}}{4\omega },  \notag \\
d &=&\frac{F_{\omega }}{\omega ^{2}}\hat{z}\cdot \vec{k},  \notag \\
e &=&\frac{F_{2\omega }F_{\omega }}{4\omega ^{3}},  \notag \\
f &=&-\frac{U_{p,1}}{2\omega },  \notag \\
g &=&-\frac{F_{2\omega }F_{\omega }}{12\omega ^{3}},  \notag
\end{eqnarray}%
and $U_{p}=U_{p,2}+U_{p,1}=(F_{2\omega }/(4\omega ))^{2}+(F_{\omega
}/(2\omega ))^{2}$ defines the ponderomotive energy as the addition of the
individual ponderomotive energies of each color. We we have dropped
diverging terms in Eq. (\ref{action2}) since only the accumulated action,
computed as differences of phases, is relevant. The two-color action of Eq. (%
\ref{action2}) reduces to the one-color action of Eq. (\ref{action-1c}) when
either $F_{\omega }=0$ (since $d=e=f=g=0$) or $F_{2\omega }=0$ (since $%
b=c=e=0$).

For the case of an $\omega -2\omega $ field with weak probe $\omega $ field
compared to the harmonic $2\omega $ field, there are four physical solutions
of Eq. (\ref{saddle}) per optical cycle (in the unit cell), i.e., $t_{1}$, $%
t_{2}$, $t_{3},$ and $t_{4}$. We have solved all the saddle times from Eq. (%
\ref{saddle}) numerically by separating their real and imaginary parts
obtaining two coupled equations:%
\begin{eqnarray}
\frac{F_{2\omega }}{2\omega }\sin \left( 2\omega \Re t_{\beta }\right)
\cosh (2\omega \Im t_{\beta }) \\ \notag
+\frac{F_{\omega }}{\omega }\sin \left(
\omega \Re t_{\beta }+\phi \right) \cosh \left( \omega \Im %
t_{\beta }\right) &=&k_{z}  \label{couple} \\
\frac{F_{2\omega }}{2\omega }\cos \left( 2\omega \Re t_{\beta }\right)
\sinh (2\omega \Im t_{\beta }) \\
+\frac{F_{\omega }}{\omega }\cos \left(
\omega \Re t_{\beta }+\phi \right) \sinh \left( \omega \Im %
t_{\beta }\right) &=&i\sqrt{2I_{p}+k_{\rho }^{2}},  \notag
\end{eqnarray}%
with $\beta =1,2,3,$ and $4.$ By neglecting either the first or the second
terms of Eq. (\ref{couple}) and summing the two equations we recover the
saddle times in Eqs. (\ref{ts0m}) and (\ref{ts0p}) for the one-color case.

\begin{figure}[tbp]
	\includegraphics[width=9cm]{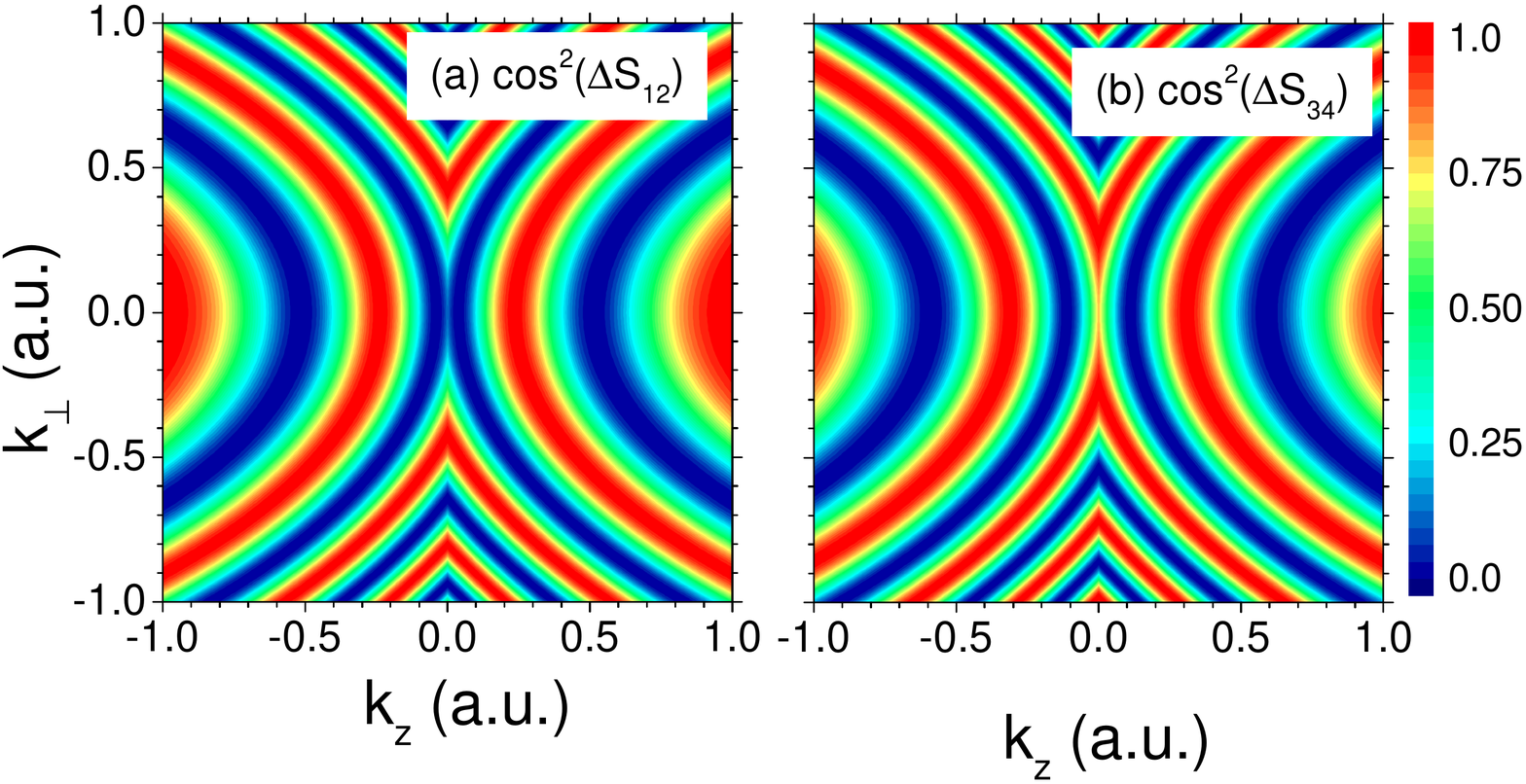}
	\caption{ Intrahalfcycle
factors $\cos ^{2}\left( \Delta S_{12}/2\right) $ in (a) and $\cos
^{2}\left( \Delta S_{34}/2\right) $ in (b).}
\label{intracycle-12-34}
\end{figure}

From Eq. (\ref{I1}), the intracycle amplitude stemming from the electron
trajectories with released times $t_{\beta }$ ($\beta =1,2,3,$ and $4$) is
proportional to%
\begin{equation}
\sum_{\beta =1}^{4}e^{iS(t_{\beta })}=e^{i\overline{S}_{1,2}}\cos \left[ 
\frac{\Delta S_{1,2}}{2}\right] +e^{i\overline{S}_{3,4}}\cos \left[ \frac{%
\Delta S_{3,4}}{2}\right]  \label{terms}
\end{equation}%
where we have omitted the prefactors of each of the terms corresponding to
the electron trajectories departing at $t_{1},t_{2},t_{3},$ and $t_{4},$ to
highlight the interference patterns. In Eq. (\ref{terms}) $\overline{S}%
_{i,j}=\left[ S(t_{i})+S(t_{j})\right] /2$ is the average action between $%
t_{i}$ and $t_{j}$ and $\Delta S_{i,j}=S(t_{j})-S(t_{i})$ is the accumulated
action between $t_{i}$ and $t_{j}.$ The accumulated actions $\Delta S_{1,2}$
and $\Delta S_{3,4}$ in the last equation contribute to the intrahalfcycle
interference of the first and second half cycles. In Fig. \ref{intracycle-12-34} we compare the
intrahalfcycle factors of the first and second half cycles. They are similar
but not equal since the vector potential in the first half cycle differs
from the second half cycle. As the probe field is weak compared to the pump
field, the intrahalfcycle distributions of Fig. \ref{intracycle-12-34} are also similar to the
one color case exhibited in Fig. \ref{1c-kzkrho}d. Taking the zeroth-order perturbation in
the probe field $\left( \Delta S_{0}\right) _{1,2}=\left( \Delta
S_{0}\right) _{3,4}\equiv \Delta S_{0},$ where $\Delta S_{0}$ denotes the
one-color accumulated action in Eq. (\ref{action-1c}) . Therefore, the
probability,calculated as the square of the absolute value of the coherent
addition of the four different terms in Eq. (\ref{terms}), can be written as%
\begin{eqnarray}
\left\vert \sum_{\beta =1}^{4}e^{iS(t_{\beta })}\right\vert ^{2} &\simeq
&\left\vert e^{i\overline{S}_{1,2}}+e^{i\overline{S}_{3,4}}\right\vert
^{2}\cos ^{2}\left[ \frac{\Delta S_{0}}{2}\right]  \notag \\
&\simeq &4\underset{\mathrm{interhalfcycle}}{\ \underbrace{\cos ^{2}\left( 
\frac{\Delta S}{2}\right) }}\underset{\mathrm{intrahalfcycle}}{\underbrace{%
\cos ^{2}\left[ \frac{\Delta S_{0}}{2}\right] }},  \label{interf}
\end{eqnarray}%
where $\Delta S=$ $\overline{S}_{3,4}-\overline{S}_{1,2}.$ To get Eq. (\ref%
{interf}) we have considered the periodicity of $S_{0}(t).$ Eq. (\ref{interf}%
) shows that the intracycle factor $\left\vert I(\vec{k})\right\vert ^{2}$
can be approximately splitted as two factors: (i) the intrahalfcycle
interference pattern $\cos ^{2}\left[ \Delta S_{0}/2\right] $ stemming from
the interference of the two electron trajectories released during half
optical cycle of the $\omega $ field (or within one optical cycle of the $%
2\omega $ field) and (ii) the interhalfcycle interference between the
contribution of the two half cycles of the $\omega $ field (or between the
two optical cycle of the $2\omega $ field).

If we go back to one-color ionization, i.e., $S(t)=S_{0}(t),$ which fulfills
the periodicity property $S_{0}(t+jT/2)=S_{0}(t)+ajT/2.$ Then, $\overline{S}%
_{3,4}=\overline{S}_{1,2}+aT/2,$ and $\Delta S=\Delta S_{0}=aT/2=a\pi
/\omega =(E+I_{p}+U_{p})\pi /\omega .$ Replacing the energy for its value at
the multiphoton peaks in Eq. (\ref{ATI1}), then $\Delta S_{0}=n\pi .$
Therefore, the intracycle factor $\cos \left( \Delta S_{0}/2\right) $ in Eq.
(\ref{interf}) becomes equal to $\pm 1$ for even $n$ (constructive
interference for the ATI peaks) and $0$ for odd $n$ (destructive
interference for the sidebands). It can be observed that the odd intercycle
rings in Fig 2a coincide with the minima of the intracycle factor in Fig.
2b. Thus, the sidebands are not formed in Fig. \ref{1c-kzkrho}c and Fig. \ref{1c-kzkrho}f, which simply
means that sidebands are only present when an $\omega $ field is applied.

\begin{figure}[tbp]
	\includegraphics[width=9cm]{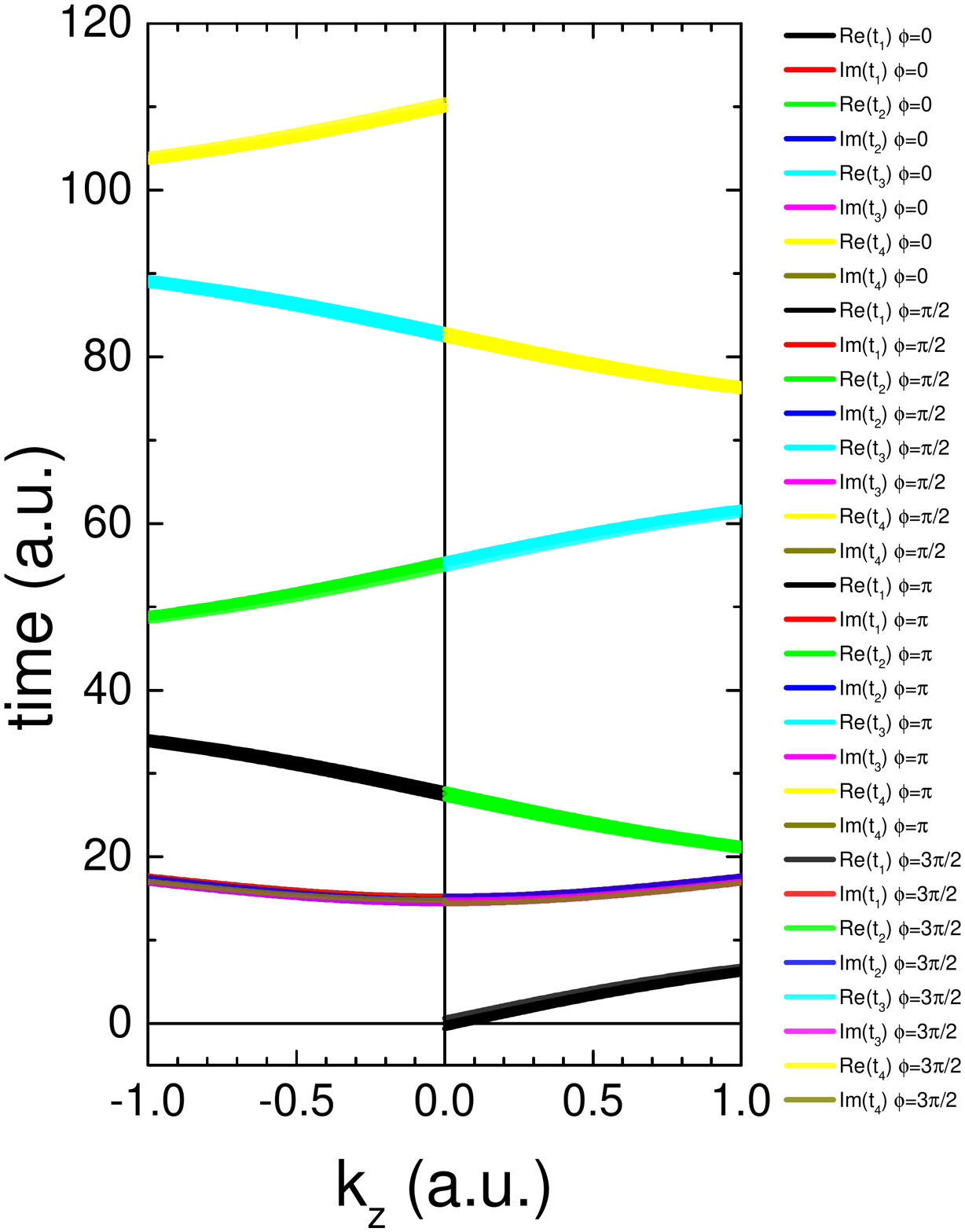}
	\caption{Complex saddle
times $t_{1}$, $t_{2}$, $t_{3}$, and $t_{4}$ as a function of the
longitudinal momentum $k_{z}$ for $k_{\bot }=0$ for $\omega -2\omega $
ionization. In solid (dash) lines the results of the SPA real (imaginary)
parts for different relative phases $\phi =0$, $\phi =\pi /2$, $\phi =\pi $,
and $\phi =3\pi /2$. The variation of both real and imaginary parts of the
saddle times with the relative phase is very small and cannot be discerned
in this figure.}
\label{saddle-times-2c}
\end{figure}

\begin{figure}[tbp]
	\includegraphics[width=10cm]{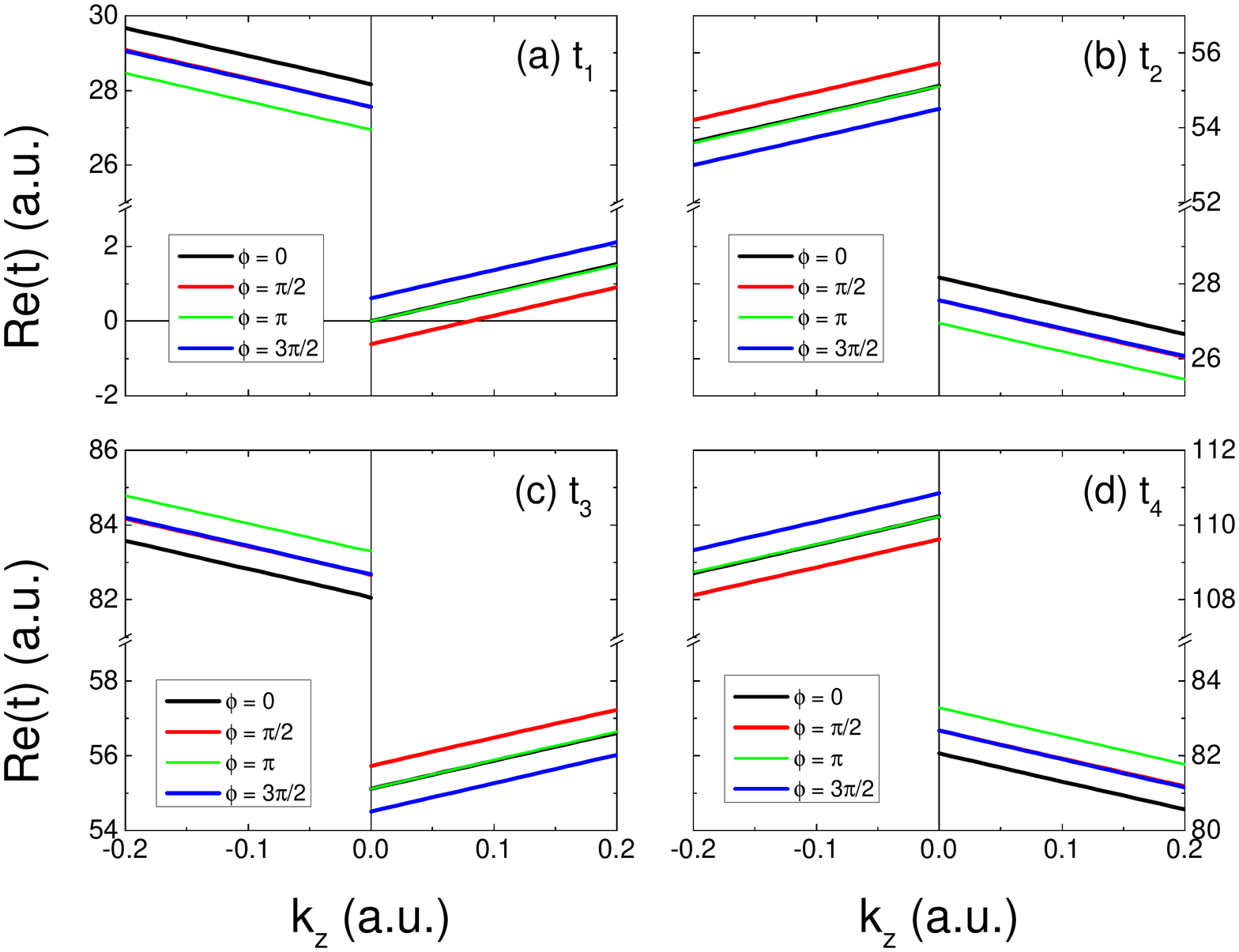}
	\caption{Close up of the real
parts of the saddle times of Fig. \ref{intracycle-12-34} for different relative phases (a) $\phi
=0$, (b) $\phi =\pi /2$, (c) $\phi =\pi $, and (d) $\phi =3\pi /2$.}
\label{closeup-t}
\end{figure}

More generally and beyond the last approximation, complex ionization
(saddle) times depend on the longitudinal and transverse momenta, similar to
the one color case. For $\omega -2\omega $ ionization, Fig. \ref{saddle-times-2c} displays the
real parts of the saddle times $t_{1}$ (in black), $t_{2}$ (in green), $%
t_{3} $ (in cyan) and $t_{4}$ (in yellow) as a function of the longitudinal
momentum $k_{z}$ for $k_{\bot }=0$ and for relative phases $\phi =0,\pi
/2,\pi ,$ and $3\pi /2.$ The same is displayed for the imaginary parts of
the ionization times. For each value of $k_{z},$ the different solutions of
Eqs. (\ref{couple}) are very similar and cannot be distinguished in the
figure. In Figs. \ref{closeup-t} we show that the apparent degeneracy with respect of the
relative phase is not such (close up of Fig. \ref{intracycle-12-34}). Whereas for $k_{z}\geq 0$,
$\Re(t_{1})$ are the same for $\phi =0$ and $\pi $ and is zero for $%
k_{z}=0,$ it has smaller values for $\phi =\pi /2$ and higher values for $%
\phi =3\pi /2$ being the difference of about $1$ a.u.. The degeneracy of $%
\Re(t_{1})$ between $\phi =0$ and $\pi $ is removed for $k_{z}<0$ at
expenses of a new degeneracy between $\phi =\pi /2$ and $3\pi /2.$ In Fig
7b, we show that $\Re(t_{2})$ lying in the second quarter cycle are
the same for $\phi =\pi /2$ and $3\pi /2$ for $k_{z}\geq 0\ $whereas it has
smaller values for $\phi =\pi $ and higher values for $\phi =0$. The
degeneracy of $\Re(t_{2})$ between $\phi =\pi /2$ and $3\pi /2$ is
removed for $k_{z}<0$ at expenses of a new degeneracy between $\phi =0$ and $%
\pi .$ $\Re(t_{3})$ lying in the third quarter cycle is shown in Figs.
6c. The degeneracy is the same as for $\Re(t_{1})$ whereas it has
smaller values for $\phi =3\pi /2$ and higher values for $\phi =\pi /2$ for $%
k_{z}\geq 0$ and the inverse for $k_{z}<0$. Finally, the same is observed
for $\Re(t_{4})$ with respect to $\Re(t_{2})$ in Figs. \ref{closeup-t}d$.$ We
do not analyze the imaginary part of the ionization times in detail since it
is not relevant for the interference patterns.

\begin{figure}[tbp]
	\includegraphics[width=9cm]{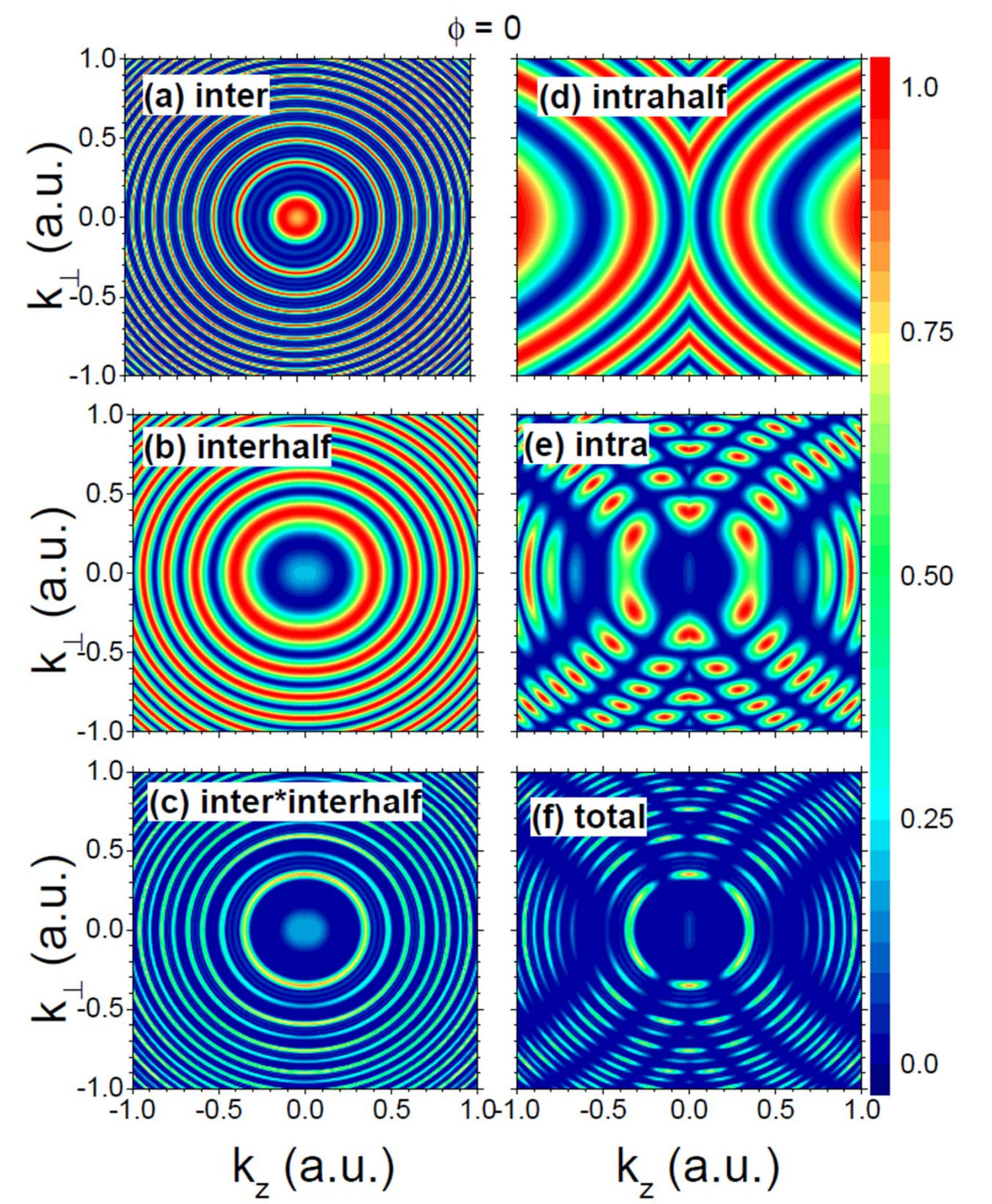}
	\caption{Doubly differential momentum distribution as a function of the longitudinal $k_{z}$
	and perpendicular momenta $k_{\bot }$ for the $\omega -2\omega $ ionization
within the SPA with relative phase $\phi =0$. (a) Intercycle factor, (b)
interhalfcycle factor, (c) multiplication of (a) and (b), (d) intrahalfcycle
factor, (e) intracycle factor [multiplication of distributions in (d) and
(b)], and (f) the total momentum distribution [multiplication of (c) and
(d)]. All distributions are normalized.}
\label{kzkrho-phi0}
\end{figure}

In Figs. \ref{kzkrho-phi0} we show the SPA doubly differential photoelectron momentum
distribution as a function of the longitudinal momentum $k_{z}$ and the
perpendicular momentum $k_{\bot }$ for zero relative phase ($\phi =0$)
between the two colors. The intercycle interference pattern for $N=4$
displays in Figs. \ref{kzkrho-phi0}a a set of multiphoton (ATI and sidebands) peaks. The
number of minima between consecutive multiphoton rings is $N-1=3.$ This
factor is practically the same as the one-color intercycle factor in Fig. \ref{1c-kzkrho}a
with an almost imperceptible difference stemming from the inclusion of the
ponderomotive energy of the $\omega $ field, which is $U_{p,1}=8.5\times
10^{-4}$ ($2\%$ of $U_{p,2}$). The intercycle factor is also independent of
the electron emission angle and the relative phase $\phi .$ The
interhalfycle factor in Figs. \ref{kzkrho-phi0}b also consists in a set of concentric rings,
but the isotropy is lost and the rings appear slightly stretched along the
longitudinal momentum. Therefore, the minima of the intracycle rings do not
perfectly match with the sidebands of the intercycle factor in Figs. \ref{kzkrho-phi0}a and
thus, they survive when one multiply the inter- and intracycle factors as
shown in Figs. \ref{kzkrho-phi0}c, unlike the one-color case. In Figs. \ref{kzkrho-phi0}d we show the
intrahalfcycle factor calculated as the intrahalfcycle of the one color
case, i.e., $\cos ^{2}\left[ \left( \Delta S_{0}\right) /2\right] $\textbf{. 
}Therefore, the intrahalfcycle pattern is independent of the relative phase $%
\phi .$ The intracycle pattern (multiplication of the interhalf- and
intrahalfcycle patterns) is shown in Figs. \ref{kzkrho-phi0}e. The quasi-isotropic
intracycle factor appears modulated by the highly angle-dependent
intrahalfcycle pattern (or viceversa). The total emission pattern is the
multiplication of the inter- (Figs. \ref{kzkrho-phi0}a) and intracycle (Figs. \ref{kzkrho-phi0}d) (see Fig.
7f). ATI peaks and sidebands of Figs. \ref{kzkrho-phi0}c modulated by the intracycle
interference pattern are observed in Figs. \ref{kzkrho-phi0}d.

\begin{figure}[tbp]
	\includegraphics[width=9cm]{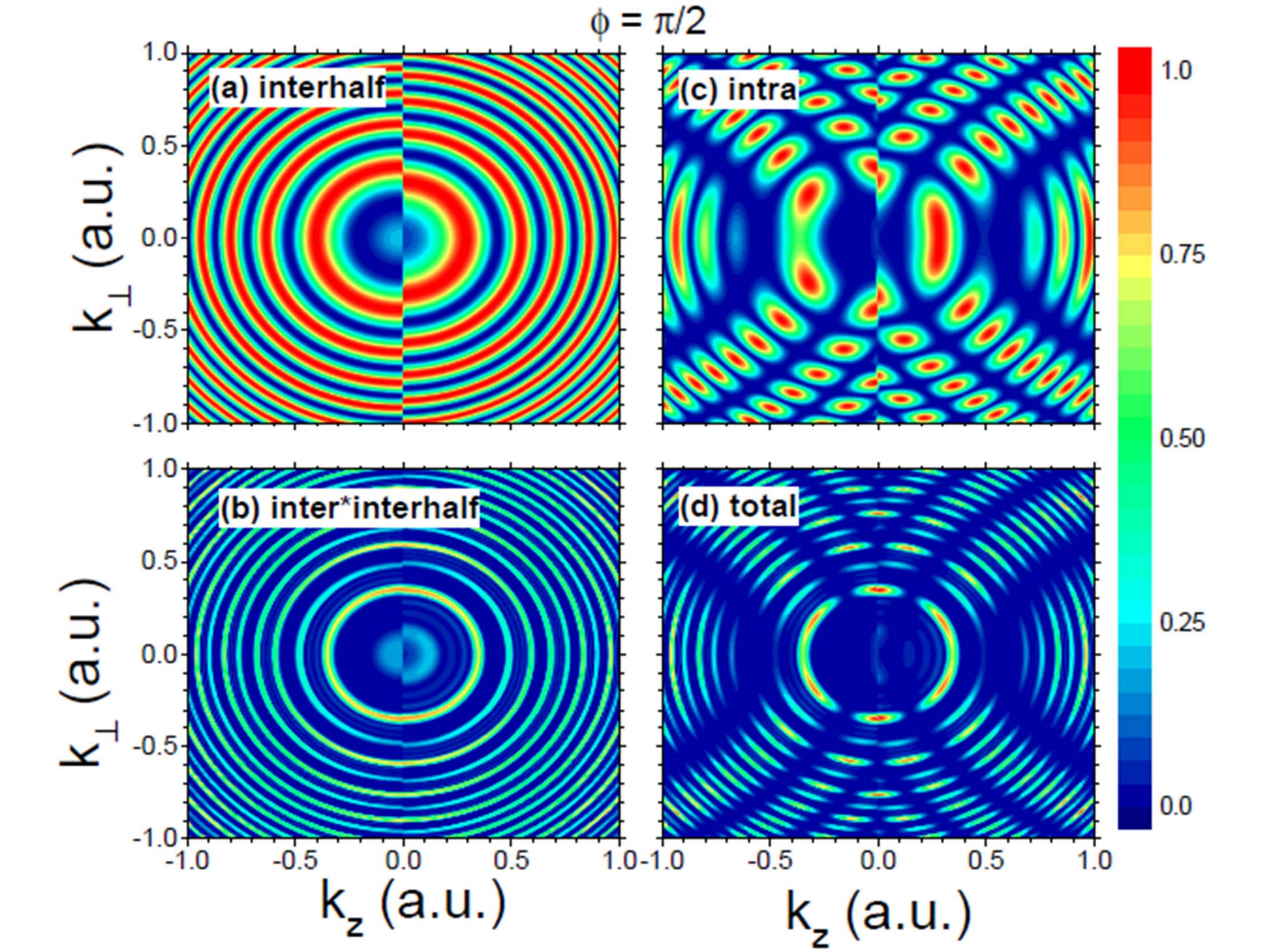}
	\caption{Doubly
differential momentum distribution as a function of the longitudinal
momentum $k_{z}$ and the perpendicular momentum $k_{\bot }$ for the $\omega
-2\omega $ ionization within the SPA with relative phase $\phi =\pi /2$. (a)
interhalfcycle factor, (b) multiplication of the intercycle factor in Fig 7a
and the interhalfcycle factor in Figs. \ref{kzkrho-phipio2}a, (c) intracycle factor
(multiplication of the intrahalfcycle factor in Figs. \ref{kzkrho-phi0}d and the
interhalfcycle of Figs. \ref{kzkrho-phipio2}a), and (d) total distribution calculated as the
multiplication of the intercycle factor in Figs. \ref{kzkrho-phi0}a and the intracycle
factor in Figs. \ref{kzkrho-phipio2}c. All distributions are normalized.}
\label{kzkrho-phipio2}
\end{figure}

The doubly differential photoelectron momentum distribution as a function of
the longitudinal momentum $k_{z}$ and the perpendicular momentum $k_{\bot }$
for relative phase $\phi =\pi /2$ is shown in Figs. \ref{kzkrho-phipio2}. The interhalfcycle
factor in Figs. \ref{kzkrho-phipio2}a also consists in a set of concentric rings with a
discontinuity for $k_{z}=0$ since the the vector potential is not
antisymmetric (with respect to the middle of the unit cell) as in the case
of $\phi =0$ (see Fig. \ref{fields-2c}b). Such discontinuities are an artifact of the SPA
and also appears for laser assisted photoionization emission (two colors
with one frequency much higher than the other) \cite{Gramajo17, Gramajo18}.
Again, the minima of the interhalfcycle rings do not match with the
sidebands of the intercycle factor in Figs. \ref{kzkrho-phi0}a and thus, they survive when
one multiply the inter- and interhalfcycle factors (see Fig. \ref{kzkrho-phipio2}b). The
intracycle factor in Fig 8c inherits the discontinuity of the interhalfcycle
factor of Figs. \ref{kzkrho-phipio2}a\textbf{. }The total emission pattern is the
multiplication of the inter- (Figs. \ref{kzkrho-phi0}a) and intracycle (Figs. \ref{kzkrho-phipio2}c) patterns
(see Fig. \ref{kzkrho-phipio2}d).

\begin{figure}[tbp]
	\includegraphics[width=9cm]{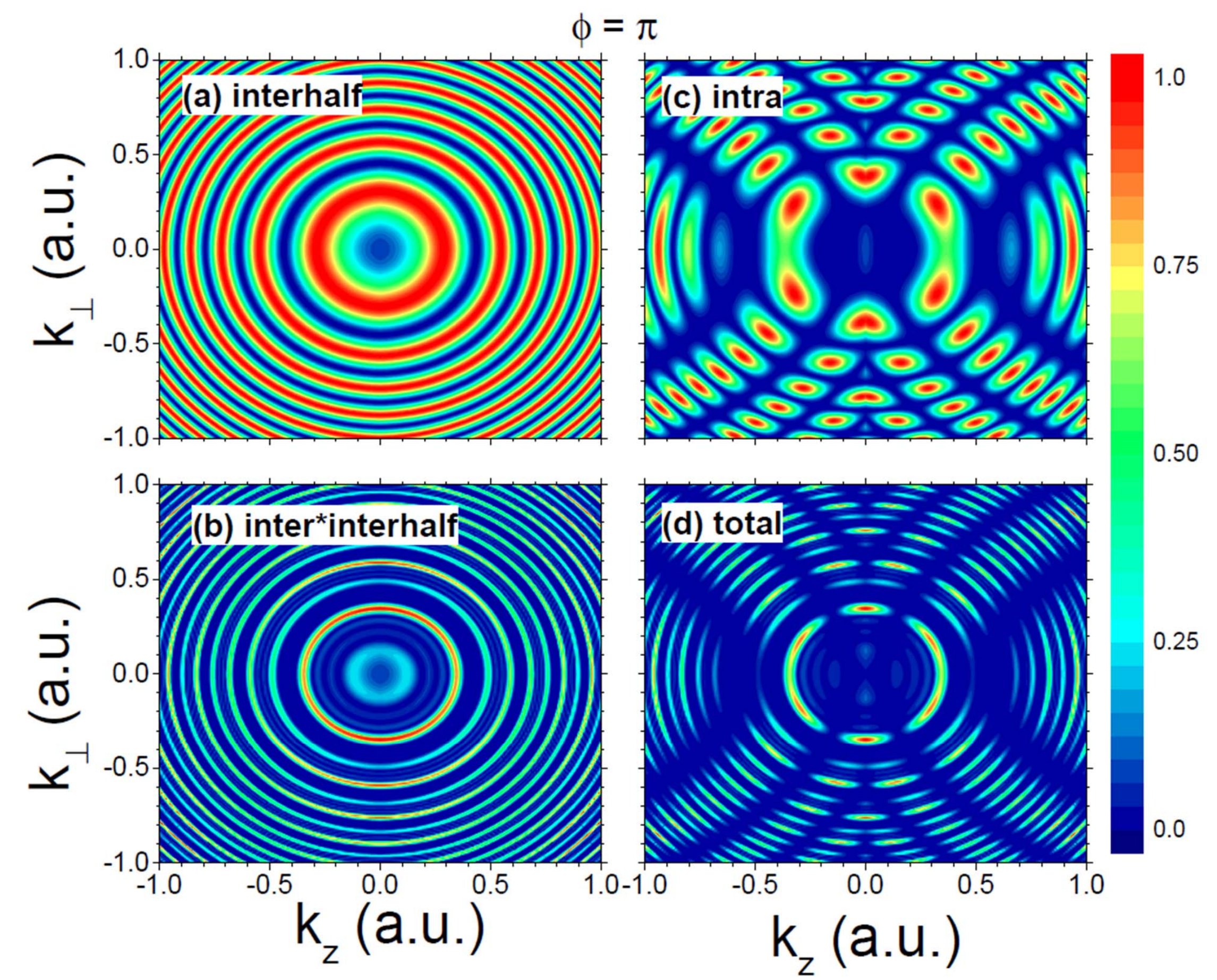}
	\caption{Doubly differential
momentum distribution as a function of the longitudinal momentum $k_{z}$ and
the perpendicular momentum $k_{\bot }$ for the $\omega -2\omega $ ionization
within the SPA with relative phase $\phi =\pi $.(a) interhalfcycle factor,
(b) multiplication of the intercycle factor in Fig 7a and the interhalfcycle
factor in Figs. \ref{kzkrho-phipi}a, (c) intracycle factor (multiplication of the
intrahalfcycle factor in Figs. \ref{kzkrho-phi0}d and the interhalfcycle of Figs. \ref{kzkrho-phipi}a), and
(d) total distribution calculated as the multiplication of the intercycle
factor in Figs. \ref{kzkrho-phi0}a and the intracycle factor in Figs. \ref{kzkrho-phipi}c. All distributions
are normalized.}
\label{kzkrho-phipi}
\end{figure}

\begin{figure}[tbp]
	\includegraphics[width=9cm]{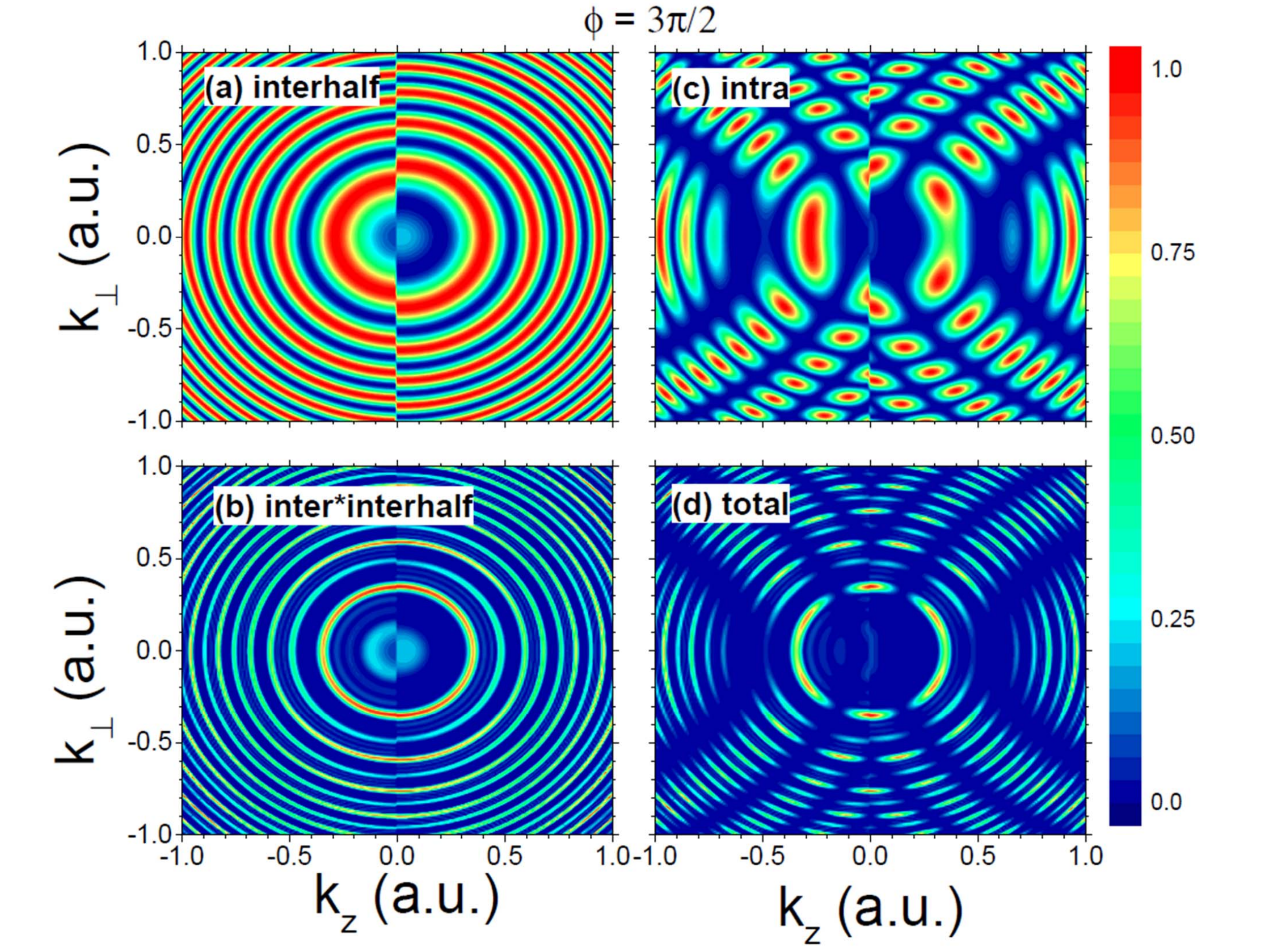}
	\caption{Doubly
differential momentum distribution as a function of the longitudinal
momentum $k_{z}$ and the perpendicular momentum $k_{\bot }$ for the $\omega
-2\omega $ ionization within the SPA with relative phase $\phi =3\pi /2$.
(a) interhalfcycle factor, (b) multiplication of the intercycle factor in
Fig 7a and the interhalfcycle factor in Fig. \ref{kzkrho-phi3pio2}a, (c) intracycle factor
(multiplication of the intrahalfcycle factor in Figs. \ref{kzkrho-phi0}d and the
interhalfcycle of Fig. \ref{kzkrho-phi3pio2}a), and (d) total distribution calculated as the
multiplication of the intercycle factor in Figs. \ref{kzkrho-phi0}a and the intracycle
factor in Fig. \ref{kzkrho-phi3pio2}c. All distributions are normalized.}
\label{kzkrho-phi3pio2}
\end{figure}

Despite the case of $\phi =\pi /2,$ for $\phi =\pi $ the two half cycles
have the same duration and the interhalfcycle distribution is continuous as
displayed in Figs. \ref{kzkrho-phipi}a (as for $\phi =0$). The multiplication of the
intercycle factor of Figs. \ref{kzkrho-phi0}a and the interhalfcycle factor of Figs. \ref{kzkrho-phipi}a is
displayed in Figs. \ref{kzkrho-phipi}b exhibiting all ATIs and sidebands. In Figs. \ref{kzkrho-phipi}c the
intracycle pattern is displayed. The total momentum distribution for $\phi
=\pi $ is shown in Figs. \ref{kzkrho-phipi}d. For $\phi =3\pi /2$ the first half cycle of the
vector potential is shorter than its second half cycle, thus, bigger
intracycle rings yield for positive $k_{z},$ as shown in Fig. \ref{kzkrho-phi3pio2}a, instead
of for negative $k_{z}$ in the case for $\phi =\pi /2.$ The multiplication
of the intercycle factor of Figs. \ref{kzkrho-phi0}a and the interhalfcycle factor of Figs.
10a is displayed in Fig. \ref{kzkrho-phi3pio2}b, which shows ATIs and sidebands. In Fig. \ref{kzkrho-phi3pio2}c
the intracycle interference pattern exhibits again a discontinuity like the
case of $\phi =\pi /2$ in Figs. \ref{kzkrho-phipio2}c. The total momentum distribution for $%
\phi =3\pi /2$ is shown in Fig. \ref{kzkrho-phi3pio2}d.

\begin{figure}[tbp]
	\includegraphics[width=9cm]{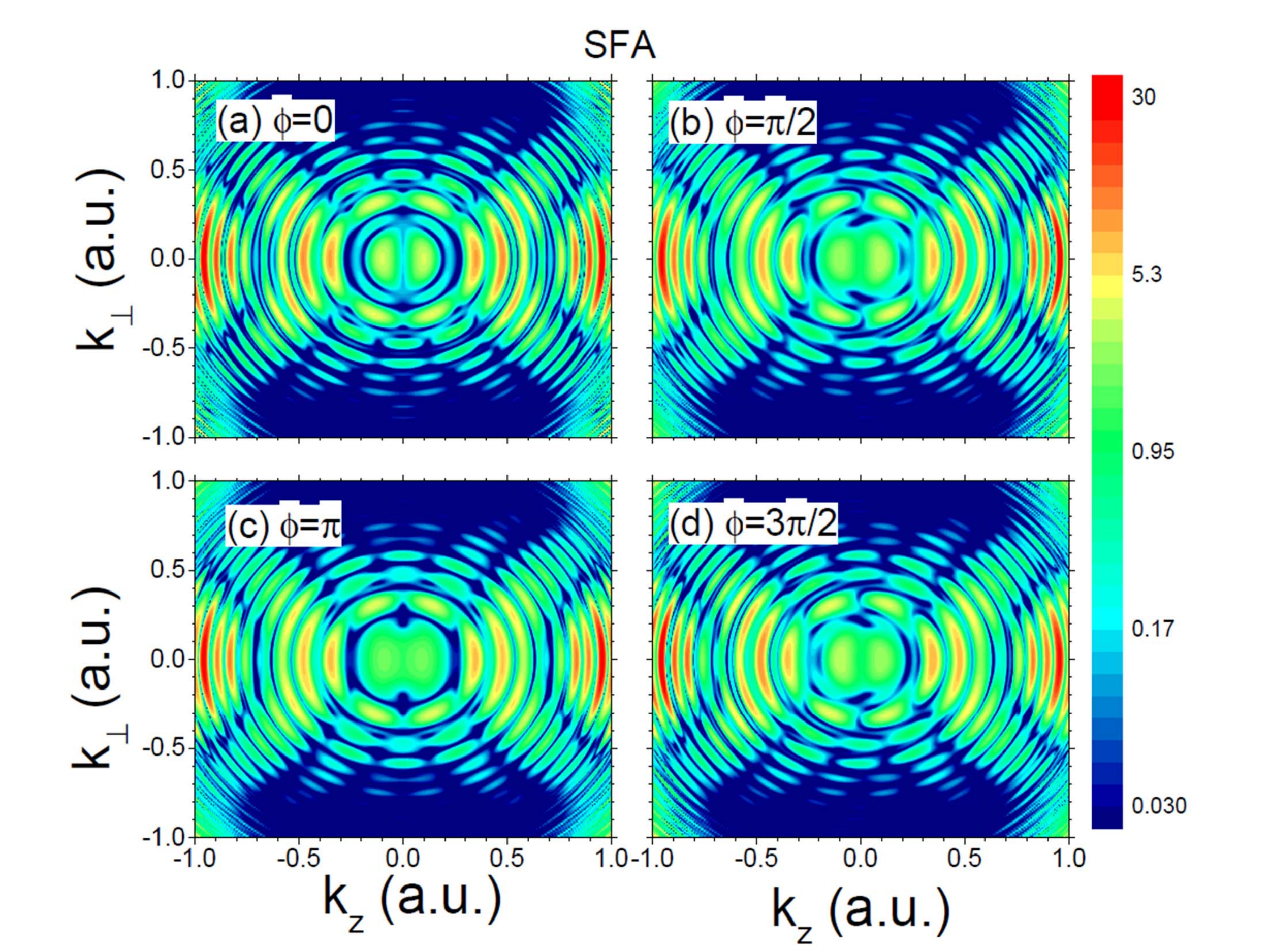}
	\caption{Doubly differential momentum
distribution as a function of the longitudinal momentum $k_{z}$ and the
perpendicular momentum $k_{\bot }$ for the $\omega -2\omega $ ionization
within the SFA with relative phase $\phi =0$ in (a), $\phi =\pi /2$ in (b), $%
\phi =\pi $ in (c), and $\phi =3\pi /2$ in (d). In order to highlight the
interference patterns we have multiplied the momentum distribution by $\exp
(10E)$ and plotted in logarithmic scale.}
\label{SFA-kzkrho-2c}
\end{figure}

In order to test the validity of the SPA, we perform the time integral in
Eq. (\ref{Tm}) numerically within the SFA \cite{Macri98,Arbo08a,Kazansky06}.
For the sake of simplicity, we model the atomic argon as a hydrogen-like
atom with effective charge $Z_{\mathrm{eff}}=\sqrt{2n^{2}I_{P}}$, where $n$
is the principal quantum number of the initial state, in this case $n=3$ and
the initial orbital quantum number is $l=1$ ($p$-state) \cite{Belkic79}.
This effective charge ensures the ionization potential to be taken into
account properly and, consequently, the intercycle fringes in the electron
spectra to be situated at the energy values of Eq. (\ref{ATI1}). We consider
an electric field with a ramp on and ramp off of duration $2\pi /\omega $
each and a flat-top region of duration $4\pi /\omega $. In Fig. \ref{Fig-anal-approx-phase} we show
the doubly differential momentum distribution for relative phase $\phi =0$
in (a), $\phi =\pi /2$ in (b), $\phi =\pi $ in (c), and $\phi =3\pi /2$ in
(d). In order to highlight the interference patterns we have multiplied the
momentum distribution by $\exp (10E)$ and plotted in logarithmic scale to
neutralize the exponential decay of the SFA as a function of the energy. The
intercycle interference pattern appeared as concentric rings situated at $%
\sqrt{2E_{n}},$ whereas the intracycle interference pattern does it with the
shape of waning and waxing moons (depending on the sign of $k_{z})$. Whereas
distributions for $\phi =0$ and $\pi $ in Figs 12a and 12c exhibit
forward-backward symmetry, the momentum distribution for $\phi =\pi /2$
results in a small asymmetry enhancing forward emission and, contrary, the
momentum distribution for $\phi =3\pi /2$ results in a small asymmetry
enhancing backward emission. The agreement between the SPA distributions in
Figs. \ref{kzkrho-phipio2}f, 9f, 10f, and 11f and the SFA distributions in Figs. \ref{SFA-kzkrho-2c} is very
good. Some differences for the angular distribution near threshold stem from
the effect of the dipole matrix element (from a $p$-state to the continuum)
in the SFA, which has been disregarded within the SPA. We have checked this
performing calculations for a hydrogenic atom from a fictitious $1s$ state
with $I_{p}=0.58$ (not shown).

\subsection{Phase delays in $\protect\omega -2\protect\omega $ ionization}

In order to get a simple close form, and considering that ionization takes
place at times near the extremes of the electric field [Eq. (\ref{field2})],
at zeroth-order perturbation, these ionization times are $t_{j}=(j-1)\pi
/(2\omega )$ with $j=1,2,3,$ and $4$. After a bit of algebra the accumulated
action becomes

\begin{eqnarray}
\Delta S &=&n\pi +\left( -d-e+g\right) \left( \cos \phi -\sin \phi \right) 
\label{DS} \\
&=&n\pi +2\chi \cos \left( \phi +\pi /4\right) ,  \notag
\end{eqnarray}%
where $\chi =\left( -d-e+g\right) /\sqrt{2}=-F_{\omega }\left( k_{z}/\omega
^{2}+F_{2\omega }/(3\omega ^{3})\right) /\sqrt{2}.$

\begin{figure}[tbp]
	\includegraphics[width=8cm]{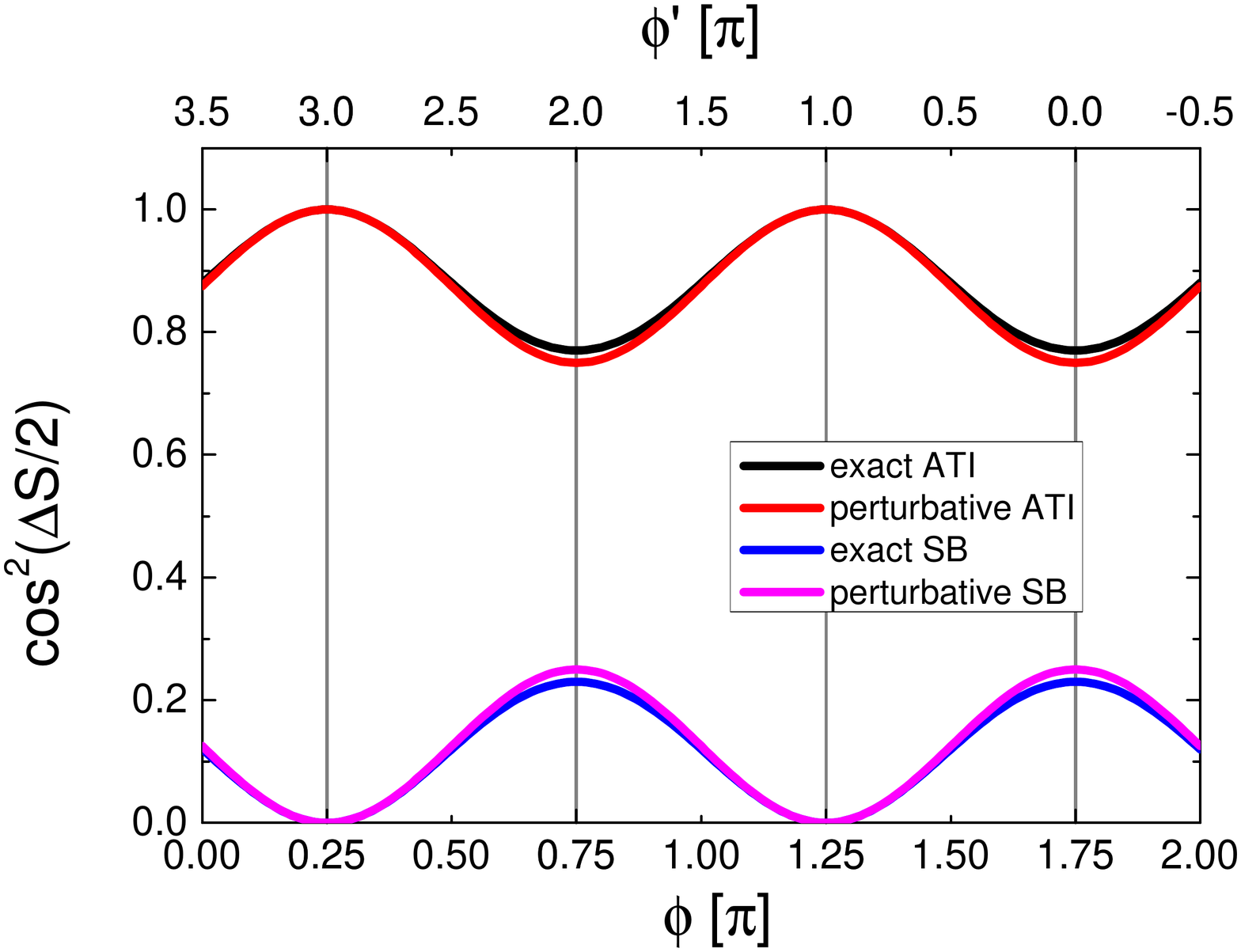}
	\caption{SPA intracycle
interference for ATI peaks and sidebands and their respective perturbative
prediction both given by Eqs. (\ref{PATI}) and (\ref{PSB}) as a function of
relative phases $\phi $ (lower axis) and $\phi ^{\prime }$ (upper axis) for $%
\chi =0.5$. ATI maximizes at $\phi =0.25\pi $ and $1.25\pi $ ($\phi ^{\prime
}=3\pi $ and $\pi $), whereas sidebands maximizes at $\phi =0.75\pi $ and $%
1.75\pi $ ($\phi ^{\prime }=2\pi $ and $0$)}
\label{Fig-anal-approx-phase}
\end{figure}

For an ATI peak, $n$ is even then replacing Eq. (\ref{DS}) into Eq. (\ref%
{interf}) we demonstrate that the intracycle interference probability factor
$\cos ^{2}\left( \frac{\Delta S}{2}\right)$ is equal to
\begin{equation}
\cos ^{2}[\chi \cos \left( \phi
+\pi /4\right) ]\simeq 1-\frac{\chi ^{2}}{2}+\frac{\chi ^{2}}{2}\cos \left[
2(\phi +3\pi /4)\right] +O(\chi ^{4}),  \label{PATI}
\end{equation}%
where in the last term we have performed a series expansion in terms of the
perturbation parameter $\chi .$ This shows that the phase delay of an ATI
peak is $\phi _{0}=-3\pi /4$. In Fig. \ref{Fig-anal-approx-phase} we display the analytical
expression $\cos ^{2}[\chi \cos \left( \phi +\pi /4\right) ]$ and its first
order perturbative approximation $1-\chi ^{2}/2+\chi ^{2}\cos \left[ 2(\phi
+3\pi /4)\right] /2$ for the value $\chi =0.5$ which reproduces the SFA
value in \cite{Zipp14} and the perturbative theory in \cite{Lopez21}.

Instead, for any sideband, $n$ is odd and thus the intracycle interference
probability factor $\cos ^{2}\left( \frac{\Delta S}{2}\right)$ can be written as%
\begin{equation}
\sin ^{2}[\chi \cos \left( \phi
+\pi /4\right) ]\simeq \frac{\chi ^{2}}{2}+\frac{\chi ^{2}}{2}\cos \left[
2(\phi +\pi /4)\right] +O(\chi ^{4}),  \label{PSB}
\end{equation}%
where in the last term we have performed a series expansion in terms of the
peturbation parameter $\chi .$ This shows that the phase delay of any
sideband is $\phi _{0}=-\pi /4,$ or equivalently\textbf{\ }$\phi
_{0}^{\prime }=-2\left( \phi _{0}+\pi /4\right) =0.$ In Fig. \ref{Fig-anal-approx-phase} we display
the analytical expression $\sin ^{2}[\chi \cos \left( \phi +\pi /4\right) ]$
and its first order perturbative approximation $1+\chi ^{2}\cos \left[
2(\phi +\pi /4)\right] /2$ for the value $\chi =0.5$ which reproduces the
SFA in \cite{Zipp14} and the perturbative theory in \cite{Lopez21}. The
variation of the probabilities of the ATI peaks and sidebands as a function
of the relative phase $\phi $ is small if $\chi $ is small, or equivalently,
if the probe field is weak. It is worth to notice from Eqs. (\ref{PATI}) and
(\ref{PSB}) that the probability of ATI peaks and sidebands as a function of
the relative phase $\phi $ have opposite phases (phase difference of $\pi $%
), as expected. The fact that the addition of the intracycle pattern of ATI
and sidebands is unity assures the conservation of probability. We obtain
similar results to the non-perturbative theory in Ref. \cite{Boll16} for
RABBIT. Especially, Eqs. (\ref{PATI}) and (\ref{PSB}) are similar to Eqs.
(14) and (15) in \cite{Boll16}. The factor $\chi \simeq -\alpha k_{z}\left(
1+F_{2\omega }/(3\omega k_{z})\right) /\sqrt{2}=-\vec{\alpha}_{\omega }\cdot 
\vec{k}\left( 1+\vec{\alpha}_{2\omega }\cdot \vec{k}/(3E_{z})\right) /\sqrt{2%
},$ where $\vec{\alpha}_{\omega }=\vec{F}_{\omega }/\omega ^{2},$ $\vec{%
\alpha}_{2\omega }=\vec{F}_{2\omega }/\left( 2\omega \right) ^{2}$
represents the quiver vector for the two different $\omega $ and $2\omega $
fields and $E_{z}=k_{z}^{2}/2$. Therefore, it may be thought that the
interhalfcycle interference pattern stem from two point sources 
\begin{eqnarray}
\cos ^{2}[\chi \cos \left( \phi +\pi /4\right) ] &=&\cos ^{2}\left[ \vec{k}%
\cdot \left( \vec{R}_{+}-\vec{R}_{-}\right) /2\right] \qquad \text{\textrm{%
(ATI)}}  \notag \\
&&  \label{molecule} \\
\sin ^{2}[\chi \cos \left( \phi +\pi /4\right) ] &=&\sin ^{2}\left[ \vec{k}%
\cdot \left( \vec{R}_{+}-\vec{R}_{-}\right) /2\right] \qquad \text{\textrm{%
(SB)}}  \notag
\end{eqnarray}%
at $\vec{R}_{+(-)}=\pm \vec{\alpha}_{\omega }\left( 1+\vec{\alpha}%
_{2\omega }\cdot \vec{k}/(3E_{z})\right) \cos \left( \phi +\pi /4\right) /%
\sqrt{2}\hat{z}$, similar to a diatomic molecule aligned along the
polarization axis. For the case of the ATIs, these two point sources emit in
phase and constructuve interference is produced at perpendicular emission;
instead, for sidebands, the two point sources emit in counterphase, leading
to partial destructive interference in the perpendicular direction since in
this case $\chi =-F_{\omega }F_{2\omega }/(3\sqrt{2}\omega ^{3})$ and not
zero as for the emission from a diatomic molecule. In Figs. \ref{kzkrho-phi0}c, 8c, 9c, and
10c we observe a minimum as a function of the angle, whereas ATIs exhibit
maxima values.

\begin{figure}[tbp]
	\includegraphics[width=9cm]{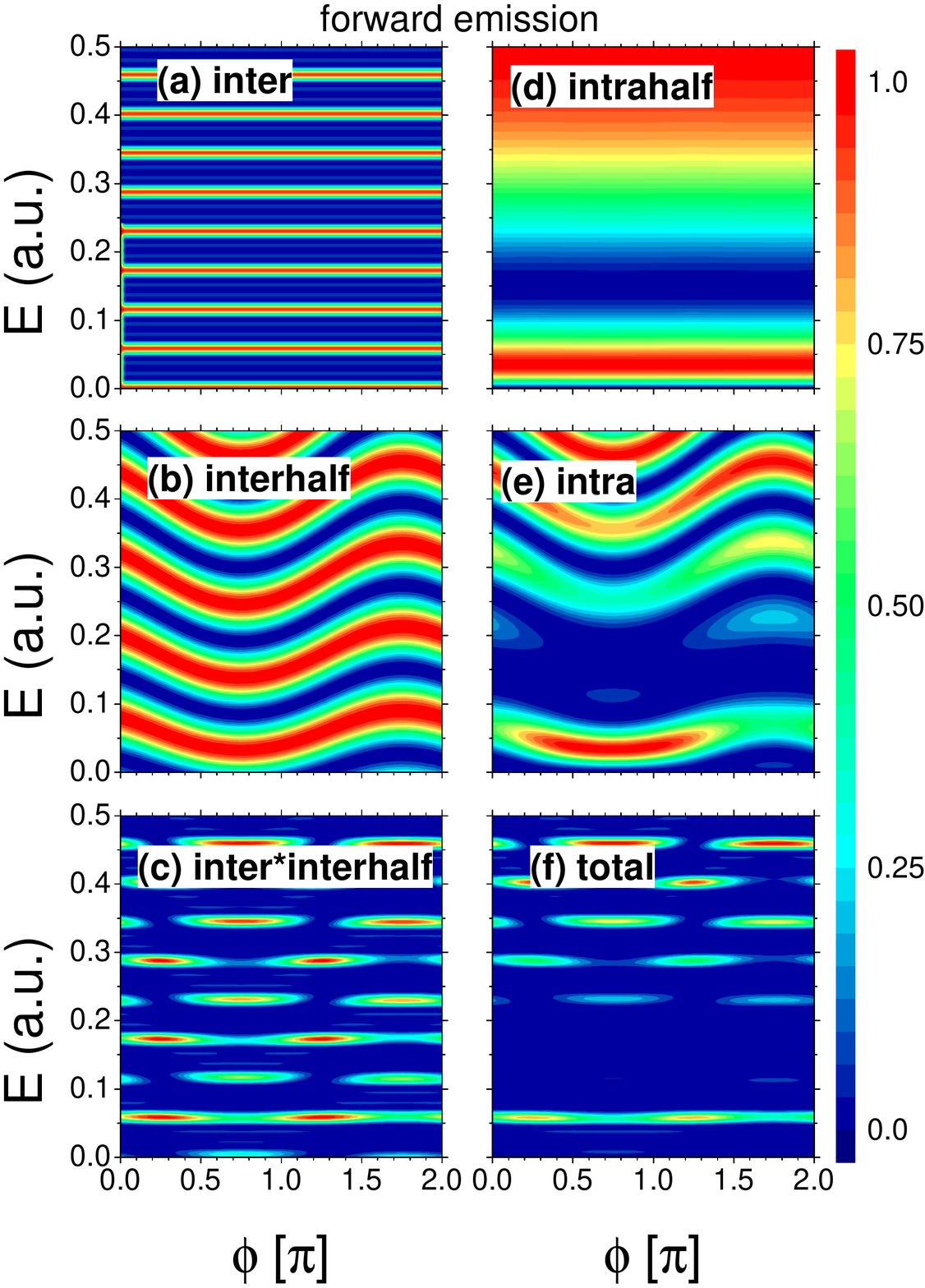}
	\caption{Energy
distribution in the forward direction as a function of the relative phase $%
\phi $ for the $\omega -2\omega $ ionization within the SPA. (a) Intercycle
factor, (b) interhalfcycle factor, (c) multiplication of (a) and (b), (d)
intrahalfcycle factor, (e) intracycle factor [multiplication of
distributions in (b) and (d)], and (f) the total momentum distribution
[multiplication of (c) and (d)]. All distributions are normalized.}
\label{forward-emission}
\end{figure}

In Fig. \ref{forward-emission} we show the energy spectrum in the forward direction as a
function of the relative phase $\phi $ between the two colors within the
SPA. In Fig 13a, we show that the intercycle factor is independent $\phi $.
All multiphoton peaks (ATIs and sidebands) are present in the intercycle
factor in Fig. \ref{forward-emission}a with separation of one $\omega $ photon energy. The
interhalfcycle factor can be observed as a $2\pi $-periodic function in Fig.
12b. The separation between interhalfcycle maxima corresponds to a $2\omega $
photon energy and the amplitude of the oscillation increases with energy,
since the accumulated action $\Delta S$ in Eq. (\ref{DS}) increases with $%
k_{z}=\sqrt{2E}$ through the factor $\chi $. The interplay between the
inter- and interhalfcycle interferences is plotted in Fig. \ref{forward-emission}c where both
ATI peaks and sidebands arise. From Eq. (\ref{ATI1}), we see that the first
multiphoton peak just above threshold ($E_{n}=0.0038$ a.u.) corresponds to $%
n=11$ and, as it is an odd number, it is a sideband with $\phi _{0}=3\pi /4$
and $7\pi /4$ (as all sidebands). In turn, ATI peaks maximize at $\phi
_{0}=\pi /4$ and $5\pi /4.$ This confirms our prediction of Eqs. (\ref{PATI}%
) and (\ref{PSB}) shown in Fig. \ref{Fig-anal-approx-phase}.

\begin{figure}[tbp]
	\includegraphics[width=9cm]{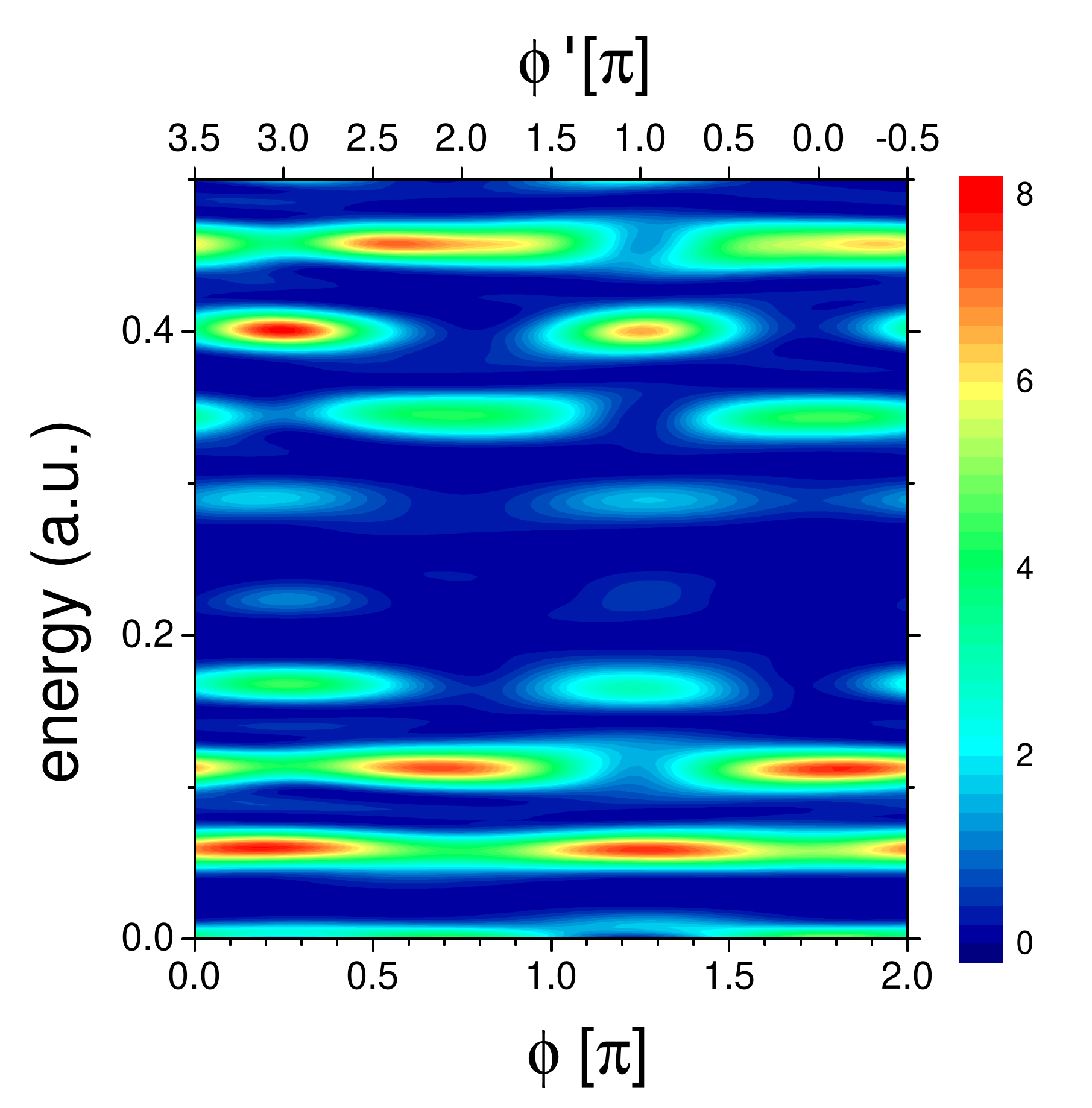}
	\caption{Energy distribution in the
forward direction as a function of the relative phase $\phi $ for the $%
\omega -2\omega $ ionization within the SFA. In order to highlight the
interference patterns we have multiplied the momentum distribution by $\exp
(15E)$}
\label{forward-SFA}
\end{figure}

We have also calculated the SFA forward emission spectrum as a function of
the relative phase $\phi $ (see Fig. \ref{forward-SFA}). In order to highlight the
interference patterns we have multiplied the momentum distribution by $\exp
(15E)$ to neutralize the exponential decay of the SFA as a function of the
energy. One observe a very good agreement between SFA (Fig. \ref{forward-SFA}) and SPA
results (Figs. \ref{forward-emission}c and 13f). We think that the small deviations of the SFA
from the SPA stem from the inclusion of the starting and ending ramps. It
can be observed a $\phi $-independent modulation with a minimum about $0.2$
a.u. similar to the intrahalfcycle interference in Fig. \ref{forward-emission}d, 13e, and 13f.
Besides,there is an anomaly in the alternation of ATIs and sidebands at
energy close to the intrahalfcycle minimum at $E\sim 0.25$ . This phenomenon
could be due to the cooper minimum stemming from the $3p$ initial state of
the argon atom.

\begin{figure}[tbp]
	\includegraphics[width=9cm]{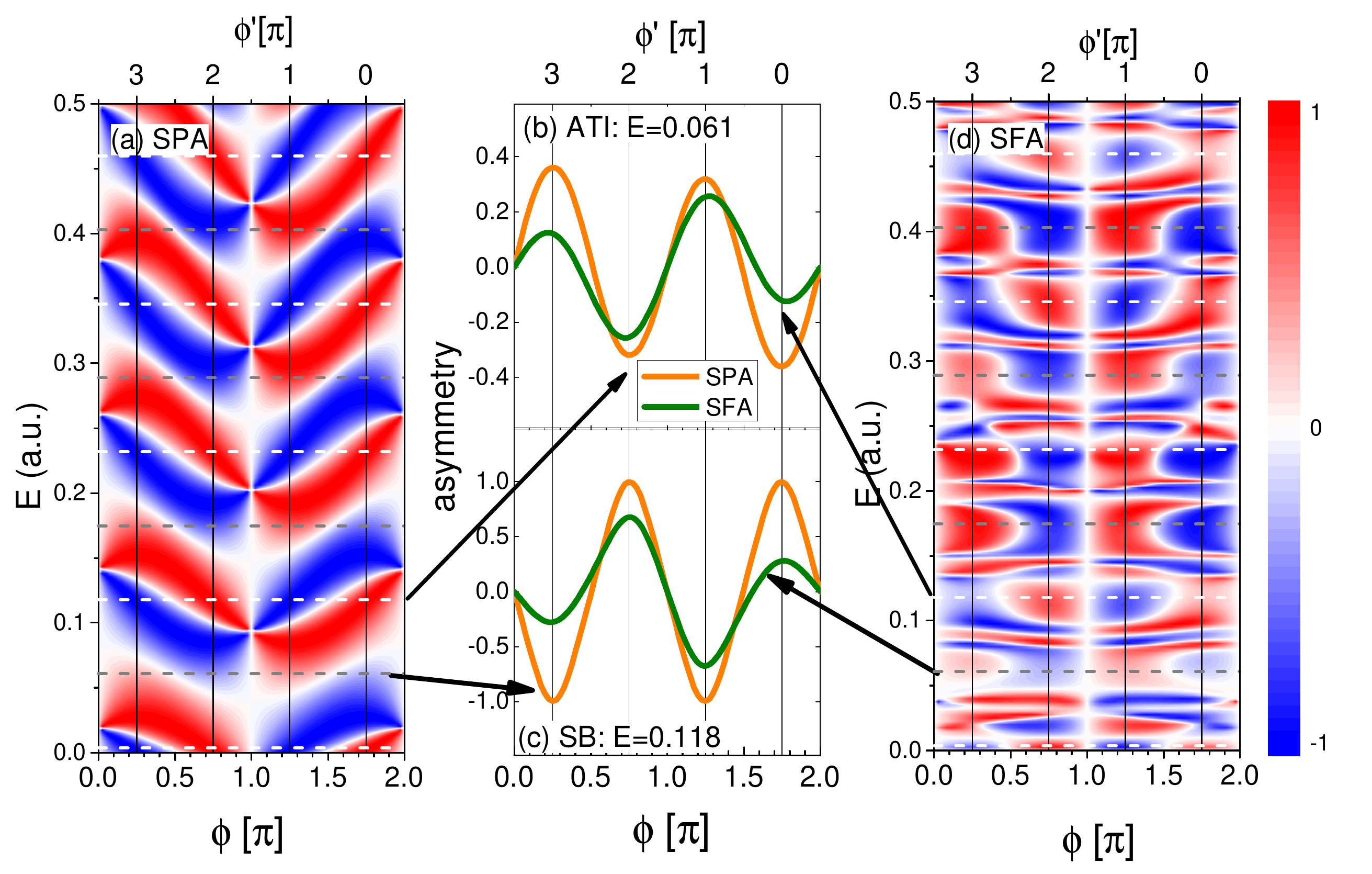}
	\caption{Asymmetry parameter as a
function of the energy $E$ and relative phase $\phi $ in (a) for the SPA and
(d) for the SFA. The horizontal dash lines indicate the energy positions of
ATIs (grey) and sidebands (white). In (b) the asymmetry parameter for the
first ATI (at $E=0.061)$ is plotted for the SPA and the SFA. In (d) the
asymmetry parameter for the second sideband (at $E=0.118)$ is plotted for
the SPA and the SFA.}
\label{symmetry}
\end{figure}

Another important quantity to map out ionization phases in the $\omega
-2\omega $ protocol is the forward-backward ($\theta \leftrightarrow \pi
-\theta $) asymmetry of the photoelectron emission probability 
\begin{equation}
A(E,\phi )=\frac{\frac{dP}{dE}(\theta =0,\phi )-\frac{dP}{dE}(\theta =\pi
,\phi )}{\frac{dP}{dE}(\theta =0,\phi )+\frac{dP}{dE}(\theta =\pi ,\phi )},
\label{asymmetry}
\end{equation}%
where the forward (backward) emission spectra $\frac{dP}{dE}(\theta =0,\phi
) $ ($\frac{dP}{dE}(\theta =\pi ,\phi )$) are defined in Eq. (\ref{dPdE}).
In Fig. \ref{asymmetry}a we show the SPA asymmetry parameter $A(E,\phi )$ as a function
of the final electron kinetic energy $E$ and the relative phase $\phi .$ The
energy positions of ATIs are marked with a horizontal grey dash line whereas
the position of sidebands are marked with a white dahline. In Fig. \ref{asymmetry}d we
show the corresponding asymmetry parameter calculated within the SFA. At
first shight, there are significant differences between the SPA and SFA
results, however, when one inspect on the asymmetry at the position of the
first ATI (in Fig. \ref{asymmetry}b) and second sideband (in Fig. \ref{asymmetry}c), similar
oscillatory behaviors are found, maximizing the ATIs at $\phi =0.25\pi $ and 
$1.25\pi $ and thee sidebands at $\phi =0.75\pi $ and $1.75\pi .$

For a close comparison between our SPA and SFA results with the perturbative
theory developed in Ref. \cite{Zipp14} accompanying an experiment for the
ionization of atomic argon by a $\omega -2\omega $ and also our recent
perturbative theory developed in Ref. \cite{Lopez21}, we perform the
transformation $t=t^{\prime }+\phi ^{\prime }/(2\omega )-\pi /(4\omega )$
and $\phi =-\phi ^{\prime }/2-\pi /4$ in Eq. (\ref{field2}), becoming the
electric field in Eq. (\ref{field1}) equivalent to the following expression
[see Eq. (1) of Ref. \cite{Zipp14} only differing in a factor 2 for the
definition of the frequencies, and Eq. (1) of Ref. \cite{Lopez21},%
\begin{equation}
\vec{F}(t^{\prime })=f(t^{\prime })\left[ F_{2\omega }\sin \left( 2\omega
t^{\prime }+\phi ^{\prime }\right) +F_{\omega }\sin (\omega t^{\prime })%
\right] \hat{z},  \label{field1}
\end{equation}%
where we have supposed that the envelope $f(t)$ remains invariant due to its
smoothness as a function of time. In Eq. (\ref{field1}) $\phi ^{\prime }$ is
the relative phase of the second harmonic with respect to the fundamental
laser field. Figs. \ref{saddle-times-2c}, 13c, 13f, and 15 show that the ATIs maximize at $\phi
_{0}=\pi /4$ and $5\pi /4,$ which is equivalent to $\phi _{0}^{\prime
}=-2\left( \phi _{0}+\pi /4\right) =\pi ,$ and $3\pi ,$ whereas the
sidebands maximize at $\phi _{0}=3\pi /4$ and $7\pi /4,$ which is equivalent
to $\phi _{0}^{\prime }=-2\left( \phi _{0}+\pi /4\right) =0,$ and $2\pi $
(modulo $2\pi $). Therefore, there are an agreement not only between our SFA
and SPA calculations but also with our own perturbation theory \cite{Lopez21}
and the perturbation theory in Ref. \cite{Zipp14}.

\section{Conclusions}

\label{conclusions}

We have developed a non-perturbative strong field theory for the atomic
ionization by a linearly polarized $\omega -2\omega $ laser pulse. We have
derived the formation of sidebands as a result of the interplay between
inter- and interhalfcycle interference patterns stemming from the effect of
a $\omega $ field with respect to a stronger $2\omega $ component. We have
individualized both interhalf- and intrahalfcycle interferences conforming
the intracycle interference pattern. We show that phase delays calculated
within our SPA agree not only with our SFA calculations but also with
previous perturbation theories \cite{Zipp14, Lopez21} extending their
validity to stronger pulses.

\medskip

\section*{Acknowledgements}

This work was supported by CONICET PIP0386, PICT-2016-0296 PICT-2017-2945
and PICT-2016-3029 of ANPCyT (Argentina). D.G.A especially thanks S. Eckart,
M. Dalhstr\"{o}m, and M. Bertolino for fruitfull discussions.

\bibliographystyle{unsrt}
\bibliography{biblio-diego}

\end{document}